\documentclass[prd,twocolumn,floatfix,amsmath,nofootinbib,amssymb,floatfix]{revtex4}
\usepackage{graphicx,color,dcolumn,booktabs,bm}
\usepackage{longtable}
\usepackage{pdflscape}
\usepackage{pdfpages}
\usepackage{txfonts}
\usepackage{overpic}
\usepackage{amssymb}
\usepackage{makecell}
\usepackage{indentfirst}
\usepackage{feynmf}   
\usepackage{slashed}  
\usepackage{cases}
\usepackage{color}
\usepackage{multirow}
\usepackage{threeparttable}
\usepackage{epstopdf}
\usepackage{enumerate}
\usepackage{ulem}
\usepackage[figuresright]{rotating}
\usepackage[colorlinks,
            citecolor=blue,
            anchorcolor=red,
            menucolor=red,
            linkcolor=red,
            filecolor=red,
            runcolor=red,
            urlcolor=blue,
            frenchlinks=true]{hyperref}

\newsavebox{\tempbox}

\begin{document}
\title{Refining radiative decay studies in singly heavy baryons}
\author{Yu-Xin Peng$^{1,2,4}$}\email{pengyx21@lzu.edu.cn} 
\author{Si-Qiang Luo$^{1,2,3,4}$}\email{luosq15@lzu.edu.cn}
\author{Xiang Liu$^{1,2,3,4}$}\email{xiangliu@lzu.edu.cn}
\affiliation{
$^1$School of Physical Science and Technology, Lanzhou University, Lanzhou 730000, China\\
$^2$Lanzhou Center for Theoretical Physics, Key Laboratory of Quantum Theory and Applications of MoE, Key Laboratory of Theoretical Physics of Gansu Province, Lanzhou University, Lanzhou 730000, China\\
$^3$MoE Frontiers Science Center for Rare Isotopes, Lanzhou University, Lanzhou 730000, China\\
$^4$Research Center for Hadron and CSR Physics, Lanzhou University $\&$ Institute of Modern Physics of CAS, Lanzhou 730000, China}

\begin{abstract}
In this work, we systematically study the radiative decays of singly heavy baryons, a crucial aspect of their spectroscopic behavior. To enhance the accuracy of our calculations, we utilize numerical spatial wave functions for the singly heavy baryons obtained through the Gaussian expansion method, which also yields their mass spectrum. As hadron spectroscopy enters an era of high precision, we believe our study of the radiative decays of singly heavy baryons will provide valuable insights for further exploration of these particles.
\end{abstract}
\maketitle

\section{Introduction}\label{sec:intro}
Recent experimental efforts have led to the discovery of numerous new hadronic states, significantly advancing the field of hadron physics over the past two decades \cite{Liu:2013waa,Cheng:2015iom,Chen:2016spr,Chen:2016qju,Klempt:2009pi,Guo:2017jvc,Brambilla:2019esw,Liu:2019zoy,Cheng:2021qpd,Meng:2022ozq,Chen:2022asf,Chu:2016sjc,Wang:2022zgi,Xu:2022kkh,Onuki:2022ugx}. Among these discoveries, singly heavy baryons stand out as a particularly interesting group. These heavy-light hadronic systems are not only numerous but have also garnered significant attention from the hadron physics community \cite{Klempt:2009pi,Cheng:2015iom,Chen:2016spr,Cheng:2021qpd,Chen:2022asf}.

A singly heavy baryon consists of one heavy quark and two light quarks, adhering well to heavy quark symmetry. This simplifies the study of certain properties by leveraging the fact that the heavy quark mass is significantly greater than the typical energy scale of strong interactions. The spectroscopy of singly heavy baryons involves analyzing their mass spectrum, decay behavior, and production processes. These studies are progressively deepening our understanding of the nonperturbative behavior of strong interactions.

Examining the over thirty single-charm baryons cataloged by the Particle Data Group (PDG)~\cite{ParticleDataGroup:2022pth}, we observe that their decay modes include not only common weak and strong decays but also radiative decays. For instance, the $\Xi_c^{\prime +}$ and $\Xi_c^{\prime 0}$ baryons, which lack Okubo-Zweig-Iizuka (OZI) allowed decay channels, predominantly decay via $\Xi_c^{+}\gamma$ and $\Xi_c^{0}\gamma$ channels, respectively, as observed by the CLEO Collaboration in 1998~\cite{CLEO:1998wvk}. Similarly, the $\Omega_c^{0}$ was first detected in the $\Omega_c^0\gamma$ channel by the BaBar Collaboration in 2006~\cite{BaBar:2006pve} and confirmed by the Belle Collaboration in 2008~\cite{Solovieva:2008fw}. In 2020, the Belle Collaboration~\cite{Belle:2020ozq} observed the radiative decays of the $\Xi_c^0(2790)$ and $\Xi_c^0(2815)$ via the $\Xi_c^0\gamma$ channel. Unlike $\Xi_c^{\prime +}$, $\Xi_c^{\prime 0}$, and $\Omega_c^*$, this marked the first observation of radiative decays of orbitally excited single-charm baryons. These findings demonstrate that radiative decays of single-charm baryons are experimentally accessible and warrant further study.

Several theoretical investigations have explored this topic. Previous works have studied radiative decays using various methods, such as light-cone QCD sum rules~\cite{Aliev:2009jt,Aliev:2014bma,Aliev:2016xvq,Wang:2009cd,Wang:2010xfj,Zhu:1998ih}, heavy quark effective theory~\cite{Cheng:1992xi,Zhu:2000py,Tawfiq:1999cf}, chiral perturbation theory~\cite{Li:2017pxa,Jiang:2015xqa,Wang:2018cre}, constituent quark models~\cite{Gandhi:2019xfw,Ortiz-Pacheco:2023kjn,Hazra:2021lpa,Wang:2023bek,Liu:2019vtx,Lu:2017meb,Xiao:2017udy,Yao:2018jmc,Wang:2017kfr,Wang:2017hej}, lattice QCD~\cite{Bahtiyar:2016dom}, relativistic quark model~\cite{Ivanov:1998wj,Ivanov:1999bk}, chiral quark-soliton model~\cite{Yang:2019tst,Kim:2021xpp}, and so on. Compared to strong decays, radiative decays exhibit different decay widths for hadrons with the same isospin but different charges. For example, theoretical calculations show that ${\rm BR}(\Xi_c^0(2790)\to \Xi_c^0\gamma)>{\rm BR}(\Xi_c^+(2790)\to \Xi_c^+\gamma)$ and ${\rm BR}(\Xi_c^0(2815)\to \Xi_c^0\gamma)>{\rm BR}(\Xi_c^+(2815)\to \Xi_c^+\gamma)$, aligning with experimental results~\cite{Belle:2020ozq}. These results indicate that radiative decays provide rich information about the internal structures of hadrons.

Hadron spectroscopy is entering an era of high precision. With the upgrade of the high-luminosity Large Hadron Collider and the operation of Belle II, more single-charm baryons will be discovered, and their properties revealed. This increase in experimental precision necessitates corresponding advancements in theoretical precision.

Given this context, our work focuses on the radiative decays of single-charm baryons, an area ripe for further exploration. Radiative decays are particularly sensitive to probing the internal structure of single-charm baryons. The calculated spatial wave function of these baryons is a crucial input for studying their radiative decays. Thus, this research provides a valuable platform for testing theoretical models.

On the other hand, over twenty single-bottom baryons have been observed in experiments. More than half of these discoveries were made from 2018, primarily due to the efforts of the LHCb and CMS Collaborations~\cite{LHCb:2018vuc,LHCb:2018haf,LHCb:2019soc,CMS:2020zzv,LHCb:2020lzx,LHCb:2020tqd,CMS:2021rvl,LHCb:2021ssn,LHCb:2023zpu}. These observations underscore the capabilities of the LHCb and CMS Collaborations in identifying single-bottom baryons. However, to date, no single-bottom baryons have been observed through radiative decays. Similar to the study of single-charm baryons, radiative decay is an effective approach for studying singly heavy baryons. For instance, there is no OZI-allowed channel for $\Omega_b^{*-}$, making $\Omega_b^- \gamma$ the dominant decay mode. Therefore, it is crucial to systematically study the radiative decays of single-bottom baryons, as this can provide different clues for their detection.

In Ref.~\cite{Luo:2023sne}, the Lanzhou group employed the Gaussian expansion method (GEM), a high-precision method for solving few-body problems~\cite{Hiyama:2003cu}, to systematically study the mass spectrum of singly heavy baryons using the potential model~\cite{Luo:2023sne,Luo:2023sra}. This approach allowed for the decoding of observed singly heavy baryon properties and the prediction of spectroscopic behavior for missing states. The spatial wave functions obtained in this process form the basis of our current work.

We revisit the radiative decays of singly heavy baryons. Previous studies~\cite{Wang:2017hej,Wang:2017kfr,Yao:2018jmc,Ortiz-Pacheco:2023kjn} often used simple harmonic oscillator wave functions to represent spatial wave functions. In contrast, we use numerical spatial wave functions calculated by GEM as presented in Ref.~\cite{Hiyama:2003cu}. The next section will compare these treatments and illustrate how they affect decay widths using a concrete example.

This timely work is in anticipation of data from Belle II~\cite{Kahn:2017ojr,Belle-II:2018jsg} and Run-3 and Run-4 at LHCb~\cite{DiNezza:2021rwt,Belin:2021jem}. Although detecting radiative decays of singly heavy baryons is more challenging than other decay modes, our study can provide theoretical guidance to experimentalists, aiding in the selection of suitable radiative decays as research topics. This will enrich observations of radiative decays of singly heavy baryons and highlight the value of our work.

This paper is organized as follows. After the Introduction, the deduction of the radiative decays of singly heavy baryons and the details of calculation are presented in Sec.~\ref{sec2}. Then the numerical results are given in Sec.~\ref{sec3}. The paper ends with a short summary in Sec.~\ref{sec4}.

\section{Radiative decays of singly heavy baryons}\label{sec2}

At the tree level, the Hamiltonian of the coupling of quarks and photon is
\begin{equation}\label{eq:He}
H_e=-\sum_{j}e_j\bar{\psi}_j\gamma_\mu^jA^\mu(\boldsymbol{k},\boldsymbol{r})\psi_j,
\end{equation}
where $e_j$, $\gamma_\mu^j$, and $\psi_j$ are the charge, Dirac matrix, and spinor of the $j$th quark, respectively. $A^\mu$ in Eq.~(\ref{eq:He}) is the photon field. In the nonrelativistic scheme, the Hamiltonian of the coupling of quarks and photons is given by~\cite{Brodsky:1968ea,Close:1970kt,Li:1994cy,Li:1997gd,Deng:2016stx,Wang:2017kfr}
\begin{equation}\label{eq:he}
h_{e}\simeq\sum_j\left[e_j \boldsymbol{r}_j\cdot \boldsymbol{\epsilon}-\frac{e_j}{2m_j}{\bm \sigma}_j\cdot(\boldsymbol{\epsilon}\times\hat{\boldsymbol{k}})\right]e^{-i\boldsymbol{ k}\cdot\boldsymbol{r}_j},
\end{equation}
where $\boldsymbol{r}_j$, $m_j$, and $\boldsymbol{\sigma}_j$ are the coordinate, mass, and Pauli matrix of the $j$th quark, respectively. $\boldsymbol{k}$ and $\boldsymbol{\epsilon}$ in Eq.~(\ref{eq:he}) are the momentum and polarization vector of the photon, respectively. With the above Hamiltonian, the radiative decay amplitude is expressed as~\cite{Deng:2016stx}
\begin{equation}\label{eq:Amp}
\mathcal{A}=-i\sqrt{\frac{\omega_\gamma}{2}}\langle f|h_e|i\rangle,
\end{equation}
where $|i\rangle$ and $|f\rangle$ are the wave functions of the initial and final baryons, respectively. The photon energy $\omega_\gamma$ is defined by
\begin{equation}\label{eq:omega}
\omega_\gamma=\frac{M_i^2-M_f^2}{2M_i}, 
\end{equation}
where $M_i$ and $M_f$ are the masses of the initial and final baryons, respectively.

Finally, we can write out the general expression of the radiative decay width of a singly heavy baryon\footnote{If we treat the quark charges as dimensionless, i.e., $e_u=\frac{2}{3}$, $e_d=-\frac{1}{3}$, $e_s=-\frac{1}{3}$, $e_c=\frac{2}{3}$, and $e_b=-\frac{1}{3}$, then Eq.~(\ref{eq:Gamma}) should be multiplied by a coefficient $4\pi\alpha_{\rm EM}$, where $\alpha_{\rm EM}\approx\frac{1}{137}$ is the fine structure constant.}
\begin{equation}\label{eq:Gamma}
\Gamma=\frac{|\boldsymbol{k}|^2}{\pi}\frac{2}{2J_i+1}\frac{M_f}{M_i}\sum_{J_{fz},J_{iz}}|\mathcal{A}_{J_{fz},J_{iz}}|^2.
\end{equation}
Here, $J_{iz}$ and $J_{fz}$ stand for the third components of the initial and final baryon angular momentum, respectively.

To obtain the radiative decay widths of singly heavy baryons, we require both their mass spectra and wave functions, which are derived from a nonrelativistic potential model~\cite{Luo:2023sne,Luo:2023sra}:
\begin{equation}\label{eq:H}
\hat{H}=\sum\limits_{i}\left(m_i+\frac{p_i^2}{2m_i}\right)+\sum\limits_{i<j}V_{ij},
\end{equation}
where $m_i$ and $p_i$ are mass and momentum of the $i$th quark, respectively. The potential $V_{ij}$ between two quarks includes the confinement potential $V_{ij}^{\rm conf}$, hyperfine interaction $V_{ij}^{\rm hyp}$, color-magnetic spin-orbit term $V_{ij}^{\rm so(cm)}$, and Thomas-precession spin-orbit term $V_{ij}^{\rm so(tp)}$: $V_{ij}=V_{ij}^{\rm conf}+V_{ij}^{\rm hyp}+V_{ij}^{\rm so(cm)}+V_{ij}^{\rm so(tp)}$. The explicit expressions for these terms are
\begin{equation}
\begin{split}
V_{ij}^{\rm conf}=&-\frac{2}{3}\frac{\alpha_s}{r_{ij}}+\frac{b}{2}r_{ij}+\frac{1}{2}C,\\
V_{ij}^{\rm hyp}=&\frac{2\alpha_s}{3m_im_j}\left[\frac{8\pi}{3}\tilde{\delta}(r_{ij}){\bf s}_i\cdot{\bf s}_j+\frac{1}{r_{ij}^3}S({\bf r},{\bf s}_i,{\bf s}_j)\right],\\
V_{ij}^{{\rm so(cm)}}=&\frac{2\alpha_s}{3r_{ij}^3}\left(\frac{{\bf r}_{ij}\times{\bf p}_i\cdot{\bf s}_i}{m_i^2}-\frac{{\bf r}_{ij}\times{\bf p}_j\cdot{\bf s}_j}{m_j^2}\right.\\
&\left.-\frac{{\bf r}_{ij}\times{\bf p}_j\cdot{\bf s}_i-{\bf r}_{ij}\times{\bf p}_i\cdot{\bf s}_j}{m_im_j}\right),\\
V_{ij}^{{\rm so(tp)}}=&-\frac{1}{2r_{ij}}\frac{\partial H_{ij}^{\rm conf}}{\partial r_{ij}}\left(\frac{{\bf r}_{ij}\times{\bf p}_i\cdot{\bf s}_i}{m_i^2}-\frac{{\bf r}_{ij}\times{\bf p}_j\cdot{\bf s}_j}{m_j^2}\right),
\end{split}
\end{equation}
where $\alpha_s$ is the coupling constant of the one-gluon exchange, $b$ is the confinement strength,  and $C$ is a mass-anchoring constant. The smearing function $\tilde{\delta}(r)=\frac{\sigma^3}{\pi^{3/2}}{\rm e}^{-\sigma^2r^2}$ is introduced in the contact term related to ${\bf s}_i\cdot {\bf s}_j$, with $\sigma$ as the smearing parameter. The tensor operator is defined as $S({\bf r},{\bf s}_i,{\bf s}_j)=\frac{3{\bf s}_i\cdot{\bf r}_{ij}{\bf s}_j\cdot{\bf r}_{ij}}{r_{ij}^2}-{\bf s}_i\cdot{\bf s}_j$.

We use the GEM~\cite{Hiyama:2003cu,Yoshida:2015tia} to solve the potential model~\cite{Luo:2023sne,Luo:2023sra} and obtain the numerical spatial wave functions of the singly heavy baryons. The Gaussian wave function is expressed as
\begin{equation}
\begin{split}
\phi_{nlm}^{\rm Gau}(\boldsymbol{r})=&R_{nl}^{\rm Gau}(r)Y_{lm}(\hat{\boldsymbol{r}}),
\end{split}
\end{equation}
where $R_{nl}^{\rm Gau}(r)$ is the radial part given by
\begin{equation}
R_{nl}^{\rm Gau}(r)=N_{nl}r^l{\rm e}^{-\nu_nr^2},~~~N_{nl}=\sqrt{\frac{2^{l+2}(2\nu_n)^{l+\frac{3}{2}}}{\sqrt{\pi}(2l+1)!!}},
\end{equation}
and $\nu_n$ is defined as
\begin{equation}
\nu_n=\frac{1}{r_n^2},\;r_n=r_1a^{n-1}\;(n=1-n_{\rm max}).
\end{equation}
In a single-channel scheme, the base could be written as
\begin{equation}\label{eq:base}
\begin{split}
|\psi_{JM}^{n_\rho n_\lambda}\rangle=\phi^{\rm color}\phi^{\rm flavor}|[[s_{q_1}s_{q_2}]_{s_\ell}[\phi_{n_\rho l_\rho}({\bm \rho}) \phi_{n_\lambda l_\lambda}({\bm \lambda})]_L]_{j_\ell}s_{Q_3}]_{JM}\rangle,
\end{split}
\end{equation}
where $\phi^{\rm color}$ and $\phi^{\rm flavor}$ are color and flavor wave functions, respectively. $s_{q_1}$, $s_{q_2}$, and $s_{Q_3}$ are spins of the quarks, with $q=u,~d,~s$, and $Q=c,~b$. $s_\ell$ and $j_\ell$ are spin and total angular momentum of the light flavor degree of freedom, respectively. A singly heavy baryon has two spatial degrees of freedom. We use the $\rho$ mode to denote the relative coordinate of the two light quarks, and the $\lambda$ mode to describe the relative position of the heavy quark and the center of mass of the two light quarks. In Eq.~(\ref{eq:base}), the $l_\rho$, $l_\lambda$, and $L$ are $\rho$-mode, $\lambda$ mode, and total orbital angular momentum, respectively. Then the spatial part in the base is expanded by
\begin{equation}
\begin{split}
[\phi_{n_\rho l_\rho}({\bm \rho}) \phi_{n_\lambda l_\lambda}({\bm \lambda})]_{LM_L}=\sum_{m_{l_\rho}m_{l_\lambda}}  &\langle l_\rho m_{l_\rho};l_\lambda m_{l_\lambda}|LM_L\rangle\\
&\times\phi_{n_\rho l_\rho m_{l_\rho}}({\bm \rho}) \phi_{n_\lambda l_\lambda m_{l_\lambda}}({\bm \lambda}).\\
\end{split}
\end{equation}
The matrix elements could be calculated with
\begin{equation}
H_{n^\prime_\rho n^\prime_\lambda n_\rho n_\lambda}=\langle\psi_{JM}^{n^\prime_\rho n^\prime_\lambda}|\hat{H}|\psi_{JM}^{n_\rho n_\lambda}\rangle,
\end{equation}
\begin{equation}
N_{n^\prime_\rho n^\prime_\lambda n_\rho n_\lambda}=\langle\psi_{JM}^{n^\prime_\rho n^\prime_\lambda}|\psi_{JM}^{n_\rho n_\lambda}\rangle.
\end{equation}
Then eigenvalues $E$ and eigenvectors $C_{n_\rho n_\lambda}$ could be obtained by the general eigenvalue equation
\begin{equation}
H_{n^\prime_\rho n^\prime_\lambda n_\rho n_\lambda}C_{n_\rho n_\lambda}=EN_{n^\prime_\rho n^\prime_\lambda n_\rho n_\lambda}C_{n_\rho n_\lambda}.
\end{equation}
Finally the total wave function is
\begin{equation}
|\Psi_{JM}\rangle=\sum\limits_{n_\rho n_\lambda} C_{n_\rho n_\lambda}|\psi_{JM}^{n_\rho n_\lambda}\rangle.
\end{equation}
In calculation, we adjust the parameters $\{r_1,r_{\rm max},n_{\rm max}\}=\{\rho_1,\rho_{\rm max},n_{\rm max}^{\rho}\}$ and $\{r_1,r_{\rm max},n_{\rm max}\}=\{\lambda_1,\lambda_{\rm max},n_{\rm max}^{\lambda}\}$ in $\rho$ and $\lambda$ modes for optimal solutions. This is the simplest situation, i.e., the quantum numbers $s_\ell$, $l_\rho$, $l_\lambda$, $L$, etc. are fixed. To be practical, we should consider the mixing effects. In this way, $s_\ell$, $l_\rho$, $l_\lambda$, $L$, etc. are also the subscripts of the matrices. And each group of the bases has independent Gaussian parameters.

It is convenient to categorize the singly heavy baryons in SU(3) flavor symmetry. In each category, the singly heavy baryons have similar heavy quark spin symmetry. In SU(3) flavor symmetry, the flavor wave functions of two light quarks can be decomposed as
\begin{equation}
3_f\otimes3_f=\bar{3}_f\oplus6_f.
\end{equation}
For singly heavy baryons, $\Lambda_c^+$, $\Xi_c^0$, $\Xi_c^+$, $\Lambda_b^0$, $\Xi_b^-$, $\Xi_b^0$ belong to the $\bar{3}_f$ representation, where the flavor wave functions of the two light quarks are antisymmetric. The symmetric flavor wave functions in the $6_f$ representation include $\Sigma_c^0$, $\Sigma_c^+$, $\Sigma_c^{++}$, $\Xi^{\prime0}_c$, $\Xi^{\prime+}_c$, $\Omega^0_c$, $\Sigma_b^-$, $\Sigma_b^0$, $\Sigma_b^{+}$, $\Xi^{\prime-}_b$, $\Xi^{\prime0}_b$, and $\Omega^-_b$.

\begin{table}[htbp]
\caption{The parameters involved in the adopted potential model.}
\label{tab:parameter}
\renewcommand\arraystretch{1.25}
\begin{tabular*}{86mm}{@{\extracolsep{\fill}}m{10mm}m{15mm}<{\centering}m{15mm}<{\centering}m{15mm}<{\centering}m{15mm}<{\centering}}\toprule[1.00pt]
\toprule[1.00pt]
System               &$\alpha_s$ &$b$ (GeV$^2$) &$\sigma$ (GeV) &$C$ (GeV)  \\
\midrule[0.75pt]
$\Lambda_c/\Sigma_c$ &0.560      &0.122         &1.600          &$-$0.633   \\
$\Xi_c^{(\prime)}$   &0.560      &0.140         &1.600          &$-$0.693   \\
$\Omega_c$           &0.578      &0.144         &1.732          &$-$0.691   \\
$\Lambda_b/\Sigma_b$ &0.560      &0.112         &1.600          &$-$0.411   \\
$\Xi_b^{(\prime)}$   &0.560      &0.123         &1.600          &$-$0.453   \\
$\Omega_b$           &0.578      &0.123         &1.732          &$-$0.435   \\
\midrule[0.75pt]
\multicolumn{5}{c}{$m_u=m_d=0.370~{\rm GeV}~~~m_{s}=0.600~{\rm GeV}$}\\
\multicolumn{5}{c}{$m_{c}=1.880~{\rm GeV}~~~m_{b}=4.977~{\rm GeV}$}\\
\bottomrule[1.00pt]
\bottomrule[1.00pt]
\end{tabular*}
\end{table}

We need consistent quark masses $m$, and potential parameters $\alpha_s$, $b$, $C$, and $\sigma$. Using all $1S$ states, $1P$, $1D$, and $2S$ $\bar{3}_f$ candidates, we determine these parameters, which are presented in Table~\ref{tab:parameter}. With these parameters, we also calculate the masses of not well-established states, such as $1F$ $\bar{3}_f$ states, $1P$, $1D$, $1F$, and $2S$ $6_f$ states. Their masses are presented in Table~\ref{tab:massspectrum}. The results match the PDG~\cite{ParticleDataGroup:2022pth} and previous studies~\cite{Roberts:2007ni,Ebert:2007bp,Ebert:2011kk,Yoshida:2015tia,Shah:2016nxi,Chen:2016iyi,Yu:2022ymb,Garcia-Tecocoatzi:2022zrf}.

\begin{table*}[htbp]
\centering
\caption{Mass spectrum of singly heavy baryons~\cite{Luo:2023sne,ParticleDataGroup:2022pth,LHCb:2018vuc,LHCb:2018haf,LHCb:2019soc,CMS:2020zzv,LHCb:2020lzx,LHCb:2020tqd,CMS:2021rvl,LHCb:2021ssn,LHCb:2023zpu}. The masses of the well-established states in the PDG~\cite{ParticleDataGroup:2022pth} or experimental results~\cite{LHCb:2018vuc,LHCb:2018haf,LHCb:2019soc,CMS:2020zzv,LHCb:2020lzx,LHCb:2020tqd,CMS:2021rvl,LHCb:2021ssn,LHCb:2023zpu} are in bold, while the unbolded masses are theoretical results. The calculated masses of single-charm and single-bottom baryons are taken from Ref.~\cite{Luo:2023sne} and the calculations of this work, respectively.}
\label{tab:massspectrum}
\renewcommand\arraystretch{1.5}
\begin{tabular*}{\textwidth}{@{\extracolsep{\fill}}cccccccccc}
\toprule[1.00pt]
\toprule[1.00pt]
&States &$l_\rho$ &$l_\lambda$ &$L$ &$s_\ell$ &$j_\ell$ &$J$ &$M~({\rm MeV})$&$M~({\rm MeV})$\\
\midrule[0.75pt]
\multirow{15}{*}{$\bar{3}_f$}&                                   &  &  &  &  &  &              &
\multirow{15}{*}{
\begin{tabular*}{0.15\textwidth}{@{\extracolsep{\fill}}cc}
$\Lambda_c$ &$\Xi_c$    \\
2283        &2472       \\
{\bf 2286}  &{\bf 2470} \\
[0.5em]
2595        &2794       \\
{\bf 2595}  &{\bf 2790} \\
[0.5em]
2618        &2817       \\
{\bf 2625}  &{\bf 2815} \\
[0.5em]
2856        &3065       \\
{\bf 2860}  &{\bf 3055} \\
[0.5em]
2867        &3075       \\
{\bf 2880}  &{\bf 3080} \\
[0.5em]
3075        &3292       \\
3079        &3295       \\
[0.5em]
2782        &2985       \\
{\bf 2765}  &{\bf 2970} \\
\end{tabular*}
}
&\multirow{15}{*}{
\begin{tabular*}{0.15\textwidth}{@{\extracolsep{\fill}}cc}
$\Lambda_b$ &$\Xi_b$    \\
5619        &5796       \\
{\bf 5619}  &{\bf 5795} \\
[0.5em]
5915        &6092       \\
{\bf 5912}  &{\bf 6087} \\
[0.5em]
5925        &6102       \\
{\bf 5920}  &{\bf 6095} \\
[0.5em]
6147        &6324       \\
{\bf 6146}  &{\bf 6327} \\
[0.5em]
6154        &6330       \\
{\bf 6152}  &{\bf 6332} \\
[0.5em]
6343        &6520       \\
6347        &6524       \\
[0.5em]
6075        &6249       \\
{\bf 6072}  &           \\
\end{tabular*}
}\\
\Xcline{9-9}{0.75pt}
\Xcline{10-10}{0.75pt}
&$\Lambda_Q/\Xi_Q(1S,\frac{1}{2}^+)$   &0 &0 &0 &0 &0 &$\frac{1}{2}$ &&\\
&                                      &  &  &  &  &  &              &&\\ 
[0.5em]
&$\Lambda_Q/\Xi_Q(1P,\frac{1}{2}^-)$   &0 &1 &1 &0 &1 &$\frac{1}{2}$ &&\\ 
&                                      &  &  &  &  &  &              &&\\ 
[0.5em]
&$\Lambda_Q/\Xi_Q(1P,\frac{3}{2}^-)$   &0 &1 &1 &0 &1 &$\frac{3}{2}$ &&\\ 
&                                      &  &  &  &  &  &              &&\\ 
[0.5em]
&$\Lambda_Q/\Xi_Q(1D,\frac{3}{2}^+)$   &0 &2 &2 &0 &2 &$\frac{3}{2}$ &&\\ 
&                                      &  &  &  &  &  &              &&\\
[0.5em]
&$\Lambda_Q/\Xi_Q(1D,\frac{5}{2}^+)$   &0 &2 &2 &0 &2 &$\frac{5}{2}$ &&\\ 
&                                      &  &  &  &  &  &              &&\\
[0.5em]
&$\Lambda_Q/\Xi_Q(1F,\frac{5}{2}^-)$   &0 &3 &3 &0 &3 &$\frac{5}{2}$ &&\\ 
&$\Lambda_Q/\Xi_Q(1F,\frac{7}{2}^-)$   &0 &3 &3 &0 &3 &$\frac{7}{2}$ &&\\
[0.5em]
&$\Lambda_Q/\Xi_Q(2S,\frac{1}{2}^+)$   &0 &0 &0 &0 &0 &$\frac{1}{2}$ &&\\
&                                      &  &  &  &  &  &              &&\\
\midrule[0.75pt]
\multirow{24}{*}{$6_f$}&                                   &  &  &  &  &  &              &
\multirow{24}{*}{
\begin{tabular*}{0.20\textwidth}{@{\extracolsep{\fill}}ccc}
$\Sigma_c$ &$\Xi_c^\prime$ &$\Omega_c$ \\
2469       &2588           &2695       \\
{\bf 2455} &{\bf 2580}     &{\bf 2695} \\
[0.5em]
2515       &2640           &2754       \\
{\bf 2520} &{\bf 2645}     &{\bf 2765} \\
[0.5em]
2785       &2919           &3030       \\
2771       &2906           &3020       \\
2798       &2938           &3056       \\
2780       &2927           &3053       \\
2797       &2945           &3073       \\
3043       &3186           &3298       \\
3058       &3201           &3315       \\
3031       &3181           &3303       \\
3043       &3194           &3317       \\
3010       &3173           &3307       \\
3017       &3179           &3314       \\
3277       &3427           &3540       \\
3283       &3433           &3546       \\
3247       &3408           &3532       \\
3252       &3412           &3536       \\
3208       &3383           &3521       \\
3209       &3383           &3521       \\
2947       &3090           &3203       \\
2979       &3124           &3243       \\
\end{tabular*}
}
&\multirow{6}{*}{
\begin{tabular*}{0.20\textwidth}{@{\extracolsep{\fill}}ccc}
$\Sigma_b$ &$\Xi_b^\prime$ &$\Omega_b$ \\
5821       &5929           &6043       \\
{\bf 5815} &{\bf 5935}     &{\bf 6046} \\
[0.5em]
5839       &5949           &6068       \\
{\bf 5835} &{\bf 5950}     &           \\
[0.5em]
6100       &6211           &6323       \\
6091       &6205           &6321       \\
6102       &6218           &6335       \\
6087       &6210           &6335       \\
6094       &6219           &6345       \\
6334       &6445           &6553       \\
6341       &6452           &6562       \\
6317       &6435           &6552       \\
6323       &6442           &6559       \\
6290       &6420           &6548       \\
6295       &6425           &6554       \\
6539       &6649           &6753       \\
6544       &6654           &6758       \\
6508       &6629           &6742       \\
6512       &6633           &6747       \\
6466       &6601           &6728       \\
6469       &6604           &6731       \\
6248       &6362           &6476       \\
6260       &6375           &6490       \\
\end{tabular*}
}\\
\Xcline{9-9}{0.75pt}
\Xcline{10-10}{0.75pt}
&$\Sigma_{Q}$\hfill$/$\hfill$\Xi^\prime_{Q}$\hfill$/$\hfill$\Omega_{Q}$\hfill$(1S,\frac{1}{2}^+)$    &0 &0 &0 &1 &1 &$\frac{1}{2}$ &\\
&                                                                                                    &  &  &  &  &  &              &\\
[0.5em]
&$\Sigma^*_{Q}$\hfill$/$\hfill$\Xi^*_{Q}$\hfill$/$\hfill$\Omega^*_{Q}$\hfill$(1S,\frac{3}{2}^+)$     &0 &0 &0 &1 &1 &$\frac{3}{2}$ &\\
&                                                                                                    &  &  &  &  &  &              &\\
[0.5em]
&$\Sigma_{Q0}/\Xi^\prime_{Q0}/\Omega_{Q0}(1P,\frac{1}{2}^-)$ &0 &1 &1 &1 &0 &$\frac{1}{2}$ &\\
&$\Sigma_{Q1}/\Xi^\prime_{Q1}/\Omega_{Q1}(1P,\frac{1}{2}^-)$ &0 &1 &1 &1 &1 &$\frac{1}{2}$ &\\
&$\Sigma_{Q1}/\Xi^\prime_{Q1}/\Omega_{Q1}(1P,\frac{3}{2}^-)$ &0 &1 &1 &1 &1 &$\frac{3}{2}$ &\\
&$\Sigma_{Q2}/\Xi^\prime_{Q2}/\Omega_{Q2}(1P,\frac{3}{2}^-)$ &0 &1 &1 &1 &2 &$\frac{3}{2}$ &\\
&$\Sigma_{Q2}/\Xi^\prime_{Q2}/\Omega_{Q2}(1P,\frac{5}{2}^-)$ &0 &1 &1 &1 &2 &$\frac{5}{2}$ &\\
&$\Sigma_{Q1}/\Xi^\prime_{Q1}/\Omega_{Q1}(1D,\frac{1}{2}^+)$ &0 &2 &2 &1 &1 &$\frac{1}{2}$ &\\
&$\Sigma_{Q1}/\Xi^\prime_{Q1}/\Omega_{Q1}(1D,\frac{3}{2}^+)$ &0 &2 &2 &1 &1 &$\frac{3}{2}$ &\\
&$\Sigma_{Q2}/\Xi^\prime_{Q2}/\Omega_{Q2}(1D,\frac{3}{2}^+)$ &0 &2 &2 &1 &2 &$\frac{3}{2}$ &\\
&$\Sigma_{Q2}/\Xi^\prime_{Q2}/\Omega_{Q2}(1D,\frac{5}{2}^+)$ &0 &2 &2 &1 &2 &$\frac{5}{2}$ &\\
&$\Sigma_{Q3}/\Xi^\prime_{Q3}/\Omega_{Q3}(1D,\frac{5}{2}^+)$ &0 &2 &2 &1 &3 &$\frac{5}{2}$ &\\
&$\Sigma_{Q3}/\Xi^\prime_{Q3}/\Omega_{Q3}(1D,\frac{7}{2}^+)$ &0 &2 &2 &1 &3 &$\frac{7}{2}$ &\\
&$\Sigma_{Q2}/\Xi^\prime_{Q2}/\Omega_{Q2}(1F,\frac{3}{2}^-)$ &0 &3 &3 &1 &2 &$\frac{3}{2}$ &\\
&$\Sigma_{Q2}/\Xi^\prime_{Q2}/\Omega_{Q2}(1F,\frac{5}{2}^-)$ &0 &3 &3 &1 &2 &$\frac{5}{2}$ &\\
&$\Sigma_{Q3}/\Xi^\prime_{Q3}/\Omega_{Q3}(1F,\frac{5}{2}^-)$ &0 &3 &3 &1 &3 &$\frac{5}{2}$ &\\
&$\Sigma_{Q3}/\Xi^\prime_{Q3}/\Omega_{Q3}(1F,\frac{7}{2}^-)$ &0 &3 &3 &1 &3 &$\frac{7}{2}$ &\\
&$\Sigma_{Q4}/\Xi^\prime_{Q4}/\Omega_{Q4}(1F,\frac{7}{2}^-)$ &0 &3 &3 &1 &4 &$\frac{7}{2}$ &\\
&$\Sigma_{Q4}/\Xi^\prime_{Q4}/\Omega_{Q4}(1F,\frac{9}{2}^-)$ &0 &3 &3 &1 &4 &$\frac{9}{2}$ &\\
&$\Sigma_{Q}$\hfill$/$\hfill$\Xi^\prime_{Q}$\hfill$/$\hfill$\Omega_{Q}$\hfill$(2S,\frac{1}{2}^+)$    &0 &0 &0 &1 &1 &$\frac{1}{2}$ &\\
&$\Sigma_{Q}^*$\hfill$/$\hfill$\Xi^*_{Q}$\hfill$/$\hfill$\Omega_{Q}^*$\hfill$(2S,\frac{3}{2}^+)$     &0 &0 &0 &1 &1 &$\frac{3}{2}$ &\\
\bottomrule[1.00pt]
\bottomrule[1.00pt]
\end{tabular*}
\end{table*}

In this work, the color, flavor, and spin wave functions are well known, and the remaining spatial wave functions are the key part, whether calculations of mass spectra or radiative decays. There are at least two approaches to obtain the spatial wave functions. The one is using the results from GEM, and the other is employing the approximations of simple harmonic oscillator (SHO) wave function \cite{Wang:2017kfr,Wang:2021hho,Luo:2023sne,Luo:2023sra}. Here, we make a comparison between the two methods. For convenience, we ignore the orbital parts, and then the remaining spatial wave functions are only two dimensional. First, with the GEM, we obtain
\begin{equation}\label{eq:Rprerholam}
R_{l_\rho l_\lambda}^{\rm GEM}(\rho,\lambda)=\sum_{n_\rho n_\lambda}C_{n_\rho n_\lambda}R_{n_\rho l_\rho }^{\rm Gau}(\rho)R_{n_\lambda l_\lambda}^{\rm Gau}(\lambda).
\end{equation}
On the other hand, with the simple harmonic oscillator wave functions, we have
\begin{equation}\label{eq:Rshorholam}
R_{n_\rho l_\rho n_\lambda l_\lambda}^{\rm SHO}(\rho,\lambda)=R_{n_\rho l_\rho }^{\rm SHO}(\beta_\rho,\rho)R_{n_\lambda l_\lambda}^{\rm SHO}(\beta_\lambda,\lambda),
\end{equation}
where $R_{nl}^{\rm SHO}(\beta,r)$ is the radial part, i.e.,
\begin{equation}
R_{nl}^{\rm SHO}(\beta,r)=\beta^{l+\frac{3}{2}}\sqrt{\frac{2n!}{\Gamma(n+l+\frac{3}{2})}}{\rm e}^{-\frac{\beta^2 r^2}{2}}L_{n}^{l+\frac{1}{2}}(\beta^2r^2)r^l.
\end{equation}
The $\beta_\rho$ and $\beta_\lambda$ are parameters to scale the SHO wave functions. In general, the $\beta_\rho$ and $\beta_\lambda$ values could not be obtained directly from the concrete potential models. However, if we know the numerical spatial wave functions from solving the potential mode, the SHO parameter $\beta_{\rho,\lambda}$ could be estimated with several different approaches. With the method suggested in Refs.~\cite{Chen:2016iyi,Chen:2018orb,Luo:2023sne,Luo:2023sra}, the $\beta_{\rho,\lambda}$ values are extracted with the spatial wave function of the GEM in Eq.~(\ref{eq:Rprerholam}), which is given by
\begin{equation}
\begin{split}
\frac{1}{\beta_\rho^2}    =&\int_0^\infty |R_{l_\rho l_\lambda}^{\rm GEM}(\rho,\lambda)|^2 \rho^4 \lambda^2 {\rm d}{\rho}{\rm d}{\lambda},\\
\frac{1}{\beta_\lambda^2} =&\int_0^\infty |R_{l_\rho l_\lambda}^{\rm GEM}(\rho,\lambda)|^2 \rho^2 \lambda^4 {\rm d}{\rho}{\rm d}{\lambda}.
\end{split}
\end{equation}
Here, we take the low-lying singly heavy baryon $\Lambda_c^+(1S,\frac{1}{2}^+)$ as an example. In the calculations, the parameters in the GEM are employed as
\begin{equation}
\begin{split}
\{\rho_1,\rho_{\rm max},n_{\rm max}^{\rho}\}=&\{0.1~{\rm fm},5.0~{\rm fm},10\},\\
\{\lambda_1,\lambda_{\rm max},n_{\rm max}^{\lambda}\}=&\{0.1~{\rm fm},5.0~{\rm fm},10\}.
\end{split}
\end{equation}
Then, using the Rayleigh-Ritz variational method,  $C_{n_{\rho},n_{\lambda}}$ is obtained as
\begin{widetext}
\begin{equation}
\renewcommand\arraystretch{1.25}
C_{n_{\rho},n_{\lambda}}=10^{-3}\times
\left(
\begin{array}{rrrrrrrrrr}
-0.093 &  0.689 & -0.687 &  2.178 &  4.442 &  4.688 & -0.876 & 0.391 &-0.155 & 0.039 \\
 2.085 & -4.407 &  9.332 & -2.680 & 33.777 & 21.436 & -2.020 & 0.612 &-0.159 & 0.024 \\
-0.973 &  9.959 &-15.248 & 27.579 & 22.564 & 36.328 & -8.591 & 4.093 &-1.698 & 0.438 \\
-1.737 & -0.326 & 32.856 &-32.591 & 89.858 & 37.447 &  0.736 &-1.416 & 0.858 &-0.268 \\
 3.663 &-12.114 &  7.243 & 94.438 & 46.207 & 87.520 &-22.094 &10.581 &-4.360 & 1.111 \\
-3.269 & 12.907 &-24.201 &  3.425 &367.804 &201.317 &-20.484 & 6.643 &-2.013 & 0.398 \\
 2.771 &-11.889 & 28.970 &-50.831 & 62.840 &203.767 &-36.996 &16.537 &-6.458 & 1.579 \\
-1.385 &  5.842 &-13.721 & 22.097 &-19.483 &-37.457 &  9.740 &-4.572 & 1.870 &-0.474 \\
 0.609 & -2.541 &  5.867 & -9.182 &  7.537 & 12.288 & -3.246 & 1.554 &-0.647 & 0.167 \\
-0.164 &  0.680 & -1.548 &  2.381 & -1.922 & -2.743 &  0.729 &-0.353 & 0.149 &-0.039 \\
\end{array}
\right),
\end{equation}
\end{widetext}
where $n_\rho$ and $n_\lambda$ are the row and column indices, respectively.\footnote{Since the singly heavy baryons discussed in this work are abundant, we do not have enough space to list all the values of $C_{n_{\rho},n_{\lambda}}$. If readers are interested in these values, they can consult Ref.~\cite{Luo:2023sne} for more details or ours.} With the coefficients obtained above, one obtains $\beta_\rho=0.290$ GeV and $\beta_\rho=0.344$ GeV in the SHO wave function [see Eq. (\ref{eq:Rshorholam})] for the $\Lambda_c^+(1S,\frac{1}{2}^+)$.

\begin{figure}[b!]
    \centering
    \includegraphics[width=8.6cm]{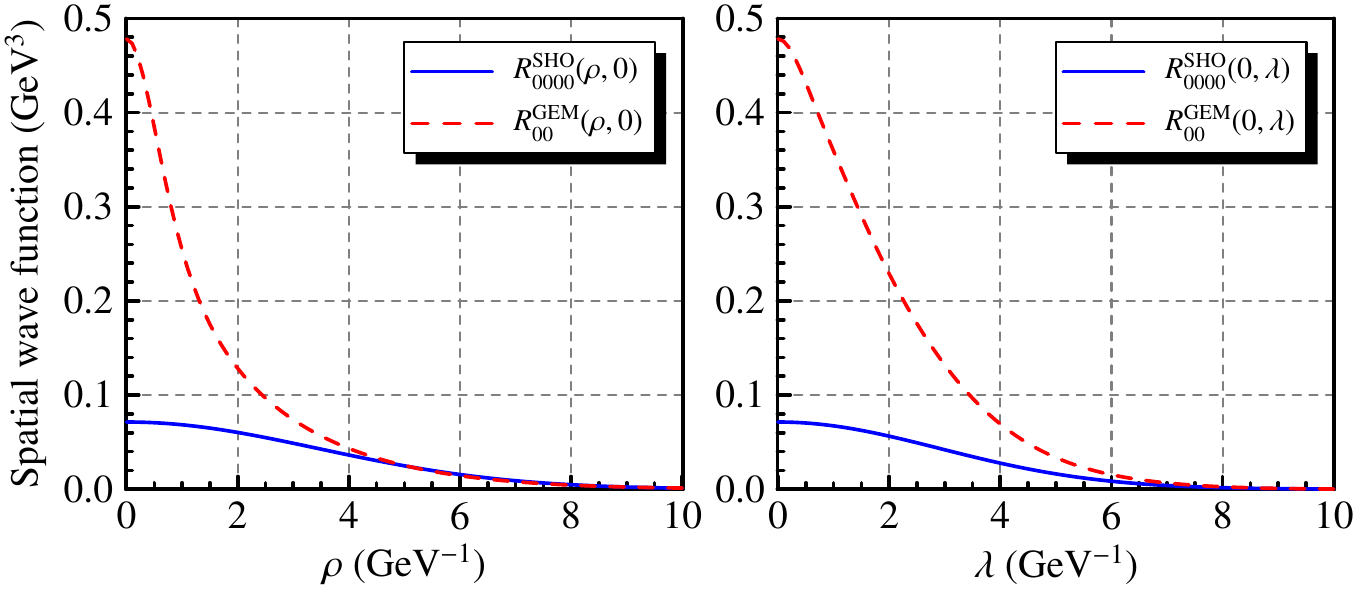}
    \caption{The spatial wave functions $R_{n_\rho l_\rho n_\lambda l_\lambda}^{\rm SHO}(\rho,\lambda)$ and $R_{l_\rho l_\lambda}^{\rm GEM}(\rho,\lambda)$ dependent on $\rho$ (left) and $\lambda$ (right).}
    \label{fig:comparewavefunction}
\end{figure}

We note that both $R_{n_\rho l_\rho n_\lambda l_\lambda}^{\rm SHO}(\rho,\lambda)$ and $R_{l_\rho l_\lambda}^{\rm GEM}(\rho,\lambda)$ are two-dimensional functions. To plot the curves conveniently, we set $\rho=0$ or $\lambda=0$ and obtain the functions depending on another variation. Here, we employ the $\Lambda_c^+(1S)$ as an example. The results are presented in Fig.~\ref{fig:comparewavefunction}. According to Fig.~\ref{fig:comparewavefunction}, there are obvious differences between the SHO and GEM wave functions. In the following discussions, we take the numerical spatial wave functions of singly heavy baryons from the potential model solved by the GEM~\cite{Luo:2023sne}, which are used as input to calculate their radiative decay behaviors.

\section{Numerical results}\label{sec3}

\subsection{$\bar{3}_f\to 1S+\gamma$}\label{subsec:bar3fto1S}

\begin{table*}[htbp]
\caption{The radiative decay widths (in units of keV) of single-charm baryons in the $\bar{3}_f$ representation decaying into the $\Lambda_c(1S)/\Sigma_c(1S)/\Xi_c^{(\prime)}(1S)\gamma$.}
\label{tab:bar3fto1S.c}
\renewcommand\arraystretch{1.25}
\centering
\begin{tabular*}{\textwidth}{@{\extracolsep{\fill}}
lcccc
lcccc
}
\toprule[1.00pt]
\toprule[1.00pt]
\multicolumn{10}{c}{
\begin{tabular*}{\textwidth}{@{\extracolsep{\fill}}
lccc
lcccc
}
 \multicolumn{1}{c}{Process}&Our& \makecell[c]{Ref.\\\cite{Wang:2017kfr}}& \makecell[c]{Ref.\\\cite{Ortiz-Pacheco:2023kjn}}
&\multicolumn{1}{c}{Process}&Our& \makecell[c]{Ref.\\\cite{Wang:2017kfr}}& \makecell[c]{Ref.\\\cite{Ortiz-Pacheco:2023kjn}}& \makecell[c]{Experiment\\\cite{Belle:2020ozq}}\\
\midrule[0.75pt]
$\Lambda_c^{+}(1P,\frac{1}{2}^-)$\hfill$\to$\hfill$\Lambda_c^{+}(1S,\frac{1}{2}^+)$\hfill$\gamma$  &0.1 &0.26 &0.1 &$\Xi_c^{0}(1P,\frac{1}{2}^-)$\hfill$\to$\hfill$\Xi_c^{0}(1S,\frac{1}{2}^+)$\hfill$\gamma$        &217.5 &263  &202.5 &$800\pm320$             \\
$\Lambda_c^{+}(1P,\frac{1}{2}^-)$\hfill$\to$\hfill$\Sigma_c^{+}(1S,\frac{1}{2}^+)$\hfill$\gamma$   &0.3 &0.45 &1.0 &$\Xi_c^{0}(1P,\frac{1}{2}^-)$\hfill$\to$\hfill$\Xi_c^{\prime 0}(1S,\frac{1}{2}^+)$\hfill$\gamma$ &  0.0 &0.0  &0.0   &$\cdots$                \\
$\Lambda_c^{+}(1P,\frac{1}{2}^-)$\hfill$\to$\hfill$\Sigma_c^{* +}(1S,\frac{3}{2}^+)$\hfill$\gamma$ &0.0 &0.05 &0.0 &$\Xi_c^{0}(1P,\frac{1}{2}^-)$\hfill$\to$\hfill$\Xi_c^{* 0}(1S,\frac{3}{2}^+)$\hfill$\gamma$      &  0.0 &0.0  &0.0   &$\cdots$                \\
$\Lambda_c^{+}(1P,\frac{3}{2}^-)$\hfill$\to$\hfill$\Lambda_c^{+}(1S,\frac{1}{2}^+)$\hfill$\gamma$  &0.8 &0.30 &0.7 &$\Xi_c^{0}(1P,\frac{3}{2}^-)$\hfill$\to$\hfill$\Xi_c^{0}(1S,\frac{1}{2}^+)$\hfill$\gamma$        &243.1 &292  &292.6 &$320\pm 45^{+45}_{-80}$ \\
$\Lambda_c^{+}(1P,\frac{3}{2}^-)$\hfill$\to$\hfill$\Sigma_c^{+}(1S,\frac{1}{2}^+)$\hfill$\gamma$   &0.9 &1.17 &2.5 &$\Xi_c^{0}(1P,\frac{3}{2}^-)$\hfill$\to$\hfill$\Xi_c^{\prime 0}(1S,\frac{1}{2}^+)$\hfill$\gamma$ &  0.0 &0.0  &0.1   &$\cdots$                \\
$\Lambda_c^{+}(1P,\frac{3}{2}^-)$\hfill$\to$\hfill$\Sigma_c^{* +}(1S,\frac{3}{2}^+)$\hfill$\gamma$ &0.2 &0.26 &0.2 &$\Xi_c^{0}(1P,\frac{3}{2}^-)$\hfill$\to$\hfill$\Xi_c^{* 0}(1S,\frac{3}{2}^+)$\hfill$\gamma$      &  0.0 &0.0  &0.0   &$\cdots$                \\
\Xcline{5-9}{0.75pt}
                                                                                                   &    &     &    &$\Xi_c^{+}(1P,\frac{1}{2}^-)$\hfill$\to$\hfill$\Xi_c^{+}(1S,\frac{1}{2}^+)$\hfill$\gamma$        &  1.7 &4.65 &7.4   &$<350$                  \\
                                                                                                   &    &     &    &$\Xi_c^{+}(1P,\frac{1}{2}^-)$\hfill$\to$\hfill$\Xi_c^{\prime +}(1S,\frac{1}{2}^+)$\hfill$\gamma$ &  1.2 &1.43 &1.3   &$\cdots$                \\
                                                                                                   &    &     &    &$\Xi_c^{+}(1P,\frac{1}{2}^-)$\hfill$\to$\hfill$\Xi_c^{* +}(1S,\frac{3}{2}^+)$\hfill$\gamma$      &  0.5 &0.44 &0.1   &$\cdots$                \\
                                                                                                   &    &     &    &$\Xi_c^{+}(1P,\frac{3}{2}^-)$\hfill$\to$\hfill$\Xi_c^{+}(1S,\frac{1}{2}^+)$\hfill$\gamma$        &  1.0 &2.8  &4.8   &$<80$                   \\
                                                                                                   &    &     &    &$\Xi_c^{+}(1P,\frac{3}{2}^-)$\hfill$\to$\hfill$\Xi_c^{\prime +}(1S,\frac{1}{2}^+)$\hfill$\gamma$ &  2.1 &2.32 &2.9   &$\cdots$                \\
                                                                                                   &    &     &    &$\Xi_c^{+}(1P,\frac{3}{2}^-)$\hfill$\to$\hfill$\Xi_c^{* +}(1S,\frac{3}{2}^+)$\hfill$\gamma$      &  1.2 &0.99 &0.3   &$\cdots$                \\
\end{tabular*}
}\\
\bottomrule[1.00pt]
\addlinespace[0.75em]
\toprule[1.00pt]
\multicolumn{10}{c}{
\begin{tabular*}{\textwidth}{@{\extracolsep{\fill}}lc@{\vrule width 0.75pt}lc@{\vrule width 0.75pt}lc}
\multicolumn{1}{c}{Process}&Our&\multicolumn{1}{c}{Process}&Our&\multicolumn{1}{c}{Process}&Our\\
\midrule[0.75pt]
$\Lambda_c^{+}(1D,\frac{3}{2}^+)$\hfill$\to$\hfill$\Lambda_c^{+}(1S,\frac{1}{2}^+)$\hfill$\gamma$  &41.3 &$\Xi_c^{0}(1D,\frac{3}{2}^+)$\hfill$\to$\hfill$\Xi_c^{0}(1S,\frac{1}{2}^+)$\hfill$\gamma$        &17.8 &$\Xi_c^{+}(1D,\frac{3}{2}^+)$\hfill$\to$\hfill$\Xi_c^{+}(1S,\frac{1}{2}^+)$\hfill$\gamma$        &28.2 \\
$\Lambda_c^{+}(1D,\frac{3}{2}^+)$\hfill$\to$\hfill$\Sigma_c^{+}(1S,\frac{1}{2}^+)$\hfill$\gamma$   & 1.4 &$\Xi_c^{0}(1D,\frac{3}{2}^+)$\hfill$\to$\hfill$\Xi_c^{\prime 0}(1S,\frac{1}{2}^+)$\hfill$\gamma$ & 0.0 &$\Xi_c^{+}(1D,\frac{3}{2}^+)$\hfill$\to$\hfill$\Xi_c^{\prime +}(1S,\frac{1}{2}^+)$\hfill$\gamma$ & 1.3 \\
$\Lambda_c^{+}(1D,\frac{3}{2}^+)$\hfill$\to$\hfill$\Sigma_c^{* +}(1S,\frac{3}{2}^+)$\hfill$\gamma$ & 1.3 &$\Xi_c^{0}(1D,\frac{3}{2}^+)$\hfill$\to$\hfill$\Xi_c^{* 0}(1S,\frac{3}{2}^+)$\hfill$\gamma$      & 0.0 &$\Xi_c^{+}(1D,\frac{3}{2}^+)$\hfill$\to$\hfill$\Xi_c^{* +}(1S,\frac{3}{2}^+)$\hfill$\gamma$      & 1.6 \\
$\Lambda_c^{+}(1D,\frac{5}{2}^+)$\hfill$\to$\hfill$\Lambda_c^{+}(1S,\frac{1}{2}^+)$\hfill$\gamma$  &42.8 &$\Xi_c^{0}(1D,\frac{5}{2}^+)$\hfill$\to$\hfill$\Xi_c^{0}(1S,\frac{1}{2}^+)$\hfill$\gamma$        &21.4 &$\Xi_c^{+}(1D,\frac{5}{2}^+)$\hfill$\to$\hfill$\Xi_c^{+}(1S,\frac{1}{2}^+)$\hfill$\gamma$        &27.1 \\
$\Lambda_c^{+}(1D,\frac{5}{2}^+)$\hfill$\to$\hfill$\Sigma_c^{+}(1S,\frac{1}{2}^+)$\hfill$\gamma$   & 1.9 &$\Xi_c^{0}(1D,\frac{5}{2}^+)$\hfill$\to$\hfill$\Xi_c^{\prime 0}(1S,\frac{1}{2}^+)$\hfill$\gamma$ & 0.0 &$\Xi_c^{+}(1D,\frac{5}{2}^+)$\hfill$\to$\hfill$\Xi_c^{\prime +}(1S,\frac{1}{2}^+)$\hfill$\gamma$ & 1.6 \\
$\Lambda_c^{+}(1D,\frac{5}{2}^+)$\hfill$\to$\hfill$\Sigma_c^{* +}(1S,\frac{3}{2}^+)$\hfill$\gamma$ & 2.0 &$\Xi_c^{0}(1D,\frac{5}{2}^+)$\hfill$\to$\hfill$\Xi_c^{* 0}(1S,\frac{3}{2}^+)$\hfill$\gamma$      & 0.0 &$\Xi_c^{+}(1D,\frac{5}{2}^+)$\hfill$\to$\hfill$\Xi_c^{* +}(1S,\frac{3}{2}^+)$\hfill$\gamma$      & 2.0 \\
$\Lambda_c^{+}(1F,\frac{5}{2}^-)$\hfill$\to$\hfill$\Lambda_c^{+}(1S,\frac{1}{2}^+)$\hfill$\gamma$  & 4.6 &$\Xi_c^{0}(1F,\frac{5}{2}^-)$\hfill$\to$\hfill$\Xi_c^{0}(1S,\frac{1}{2}^+)$\hfill$\gamma$        &28.4 &$\Xi_c^{+}(1F,\frac{5}{2}^-)$\hfill$\to$\hfill$\Xi_c^{+}(1S,\frac{1}{2}^+)$\hfill$\gamma$        & 0.1 \\
$\Lambda_c^{+}(1F,\frac{5}{2}^-)$\hfill$\to$\hfill$\Sigma_c^{+}(1S,\frac{1}{2}^+)$\hfill$\gamma$   & 0.9 &$\Xi_c^{0}(1F,\frac{5}{2}^-)$\hfill$\to$\hfill$\Xi_c^{\prime 0}(1S,\frac{1}{2}^+)$\hfill$\gamma$ & 0.0 &$\Xi_c^{+}(1F,\frac{5}{2}^-)$\hfill$\to$\hfill$\Xi_c^{\prime +}(1S,\frac{1}{2}^+)$\hfill$\gamma$ & 0.6 \\
$\Lambda_c^{+}(1F,\frac{5}{2}^-)$\hfill$\to$\hfill$\Sigma_c^{* +}(1S,\frac{3}{2}^+)$\hfill$\gamma$ & 1.2 &$\Xi_c^{0}(1F,\frac{5}{2}^-)$\hfill$\to$\hfill$\Xi_c^{* 0}(1S,\frac{3}{2}^+)$\hfill$\gamma$      & 0.0 &$\Xi_c^{+}(1F,\frac{5}{2}^-)$\hfill$\to$\hfill$\Xi_c^{* +}(1S,\frac{3}{2}^+)$\hfill$\gamma$      & 0.9 \\
$\Lambda_c^{+}(1F,\frac{7}{2}^-)$\hfill$\to$\hfill$\Lambda_c^{+}(1S,\frac{1}{2}^+)$\hfill$\gamma$  & 4.8 &$\Xi_c^{0}(1F,\frac{7}{2}^-)$\hfill$\to$\hfill$\Xi_c^{0}(1S,\frac{1}{2}^+)$\hfill$\gamma$        &26.9 &$\Xi_c^{+}(1F,\frac{7}{2}^-)$\hfill$\to$\hfill$\Xi_c^{+}(1S,\frac{1}{2}^+)$\hfill$\gamma$        & 0.1 \\
$\Lambda_c^{+}(1F,\frac{7}{2}^-)$\hfill$\to$\hfill$\Sigma_c^{+}(1S,\frac{1}{2}^+)$\hfill$\gamma$   & 0.9 &$\Xi_c^{0}(1F,\frac{7}{2}^-)$\hfill$\to$\hfill$\Xi_c^{\prime 0}(1S,\frac{1}{2}^+)$\hfill$\gamma$ & 0.0 &$\Xi_c^{+}(1F,\frac{7}{2}^-)$\hfill$\to$\hfill$\Xi_c^{\prime +}(1S,\frac{1}{2}^+)$\hfill$\gamma$ & 0.6 \\
$\Lambda_c^{+}(1F,\frac{7}{2}^-)$\hfill$\to$\hfill$\Sigma_c^{* +}(1S,\frac{3}{2}^+)$\hfill$\gamma$ & 1.2 &$\Xi_c^{0}(1F,\frac{7}{2}^-)$\hfill$\to$\hfill$\Xi_c^{* 0}(1S,\frac{3}{2}^+)$\hfill$\gamma$      & 0.0 &$\Xi_c^{+}(1F,\frac{7}{2}^-)$\hfill$\to$\hfill$\Xi_c^{* +}(1S,\frac{3}{2}^+)$\hfill$\gamma$      & 0.9 \\
$\Lambda_c^{+}(2S,\frac{1}{2}^+)$\hfill$\to$\hfill$\Lambda_c^{+}(1S,\frac{1}{2}^+)$\hfill$\gamma$  & 0.0 &$\Xi_c^{0}(2S,\frac{1}{2}^+)$\hfill$\to$\hfill$\Xi_c^{0}(1S,\frac{1}{2}^+)$\hfill$\gamma$        & 0.0 &$\Xi_c^{+}(2S,\frac{1}{2}^+)$\hfill$\to$\hfill$\Xi_c^{+}(1S,\frac{1}{2}^+)$\hfill$\gamma$        & 0.0 \\
$\Lambda_c^{+}(2S,\frac{1}{2}^+)$\hfill$\to$\hfill$\Sigma_c^{+}(1S,\frac{1}{2}^+)$\hfill$\gamma$   & 0.0 &$\Xi_c^{0}(2S,\frac{1}{2}^+)$\hfill$\to$\hfill$\Xi_c^{\prime 0}(1S,\frac{1}{2}^+)$\hfill$\gamma$ & 0.1 &$\Xi_c^{+}(2S,\frac{1}{2}^+)$\hfill$\to$\hfill$\Xi_c^{\prime +}(1S,\frac{1}{2}^+)$\hfill$\gamma$ & 0.9 \\
$\Lambda_c^{+}(2S,\frac{1}{2}^+)$\hfill$\to$\hfill$\Sigma_c^{* +}(1S,\frac{3}{2}^+)$\hfill$\gamma$ & 3.2 &$\Xi_c^{0}(2S,\frac{1}{2}^+)$\hfill$\to$\hfill$\Xi_c^{* 0}(1S,\frac{3}{2}^+)$\hfill$\gamma$      & 0.0 &$\Xi_c^{+}(2S,\frac{1}{2}^+)$\hfill$\to$\hfill$\Xi_c^{* +}(1S,\frac{3}{2}^+)$\hfill$\gamma$      & 1.4 \\
\end{tabular*}
}\\
\bottomrule[1.00pt]
\bottomrule[1.00pt]
\end{tabular*}
\end{table*}

\begin{table*}[htbp]
\caption{The radiative decay widths (in units of keV) of single-bottom baryons in the $\bar{3}_f$ representation decaying into the $\Lambda_b(1S)/\Sigma_b(1S)/\Xi_b^{(\prime)}(1S)\gamma$.}
\label{tab:bar3fto1S.b}
\renewcommand\arraystretch{1.25}
\centering
\begin{tabular*}{\textwidth}{@{\extracolsep{\fill}}
lcccc
lcccc
}
\toprule[1.00pt]
\toprule[1.00pt]
\multicolumn{10}{c}{
\begin{tabular*}{\textwidth}{@{\extracolsep{\fill}}
lccc
lccc
}
 \multicolumn{1}{c}{Process}&Our& \makecell[c]{Ref.\\\cite{Wang:2017kfr}}& \makecell[c]{Ref.\\\cite{Ortiz-Pacheco:2023kjn}}
&\multicolumn{1}{c}{Process}&Our& \makecell[c]{Ref.\\\cite{Wang:2017kfr}}& \makecell[c]{Ref.\\\cite{Ortiz-Pacheco:2023kjn}}\\
\midrule[0.75pt]
$\Lambda_b^{0}(1P,\frac{1}{2}^-)$\hfill$\to$\hfill$\Lambda_b^{0}(1S,\frac{1}{2}^+)$\hfill$\gamma$  &47.0 &50.2 &40.7 &$\Xi_b^{-}(1P,\frac{1}{2}^-)$\hfill$\to$\hfill$\Xi_b^{-}(1S,\frac{1}{2}^+)$\hfill$\gamma$        &79.1 &135  &91.5 \\
$\Lambda_b^{0}(1P,\frac{1}{2}^-)$\hfill$\to$\hfill$\Sigma_b^{0}(1S,\frac{1}{2}^+)$\hfill$\gamma$   & 0.1 &0.14 &0.2  &$\Xi_b^{-}(1P,\frac{1}{2}^-)$\hfill$\to$\hfill$\Xi_b^{\prime -}(1S,\frac{1}{2}^+)$\hfill$\gamma$ & 0.0 &0.0  &0.0  \\
$\Lambda_b^{0}(1P,\frac{1}{2}^-)$\hfill$\to$\hfill$\Sigma_b^{* 0}(1S,\frac{3}{2}^+)$\hfill$\gamma$ & 0.1 &0.09 &0.0  &$\Xi_b^{-}(1P,\frac{1}{2}^-)$\hfill$\to$\hfill$\Xi_b^{* -}(1S,\frac{3}{2}^+)$\hfill$\gamma$      & 0.0 &0.0  &0.0  \\
$\Lambda_b^{0}(1P,\frac{3}{2}^-)$\hfill$\to$\hfill$\Lambda_b^{0}(1S,\frac{1}{2}^+)$\hfill$\gamma$  &49.1 &52.8 &43.4 &$\Xi_b^{-}(1P,\frac{3}{2}^-)$\hfill$\to$\hfill$\Xi_b^{-}(1S,\frac{1}{2}^+)$\hfill$\gamma$        &84.5 &147  &96.1 \\
$\Lambda_b^{0}(1P,\frac{3}{2}^-)$\hfill$\to$\hfill$\Sigma_b^{0}(1S,\frac{1}{2}^+)$\hfill$\gamma$   & 0.1 &0.21 &0.3  &$\Xi_b^{-}(1P,\frac{3}{2}^-)$\hfill$\to$\hfill$\Xi_b^{\prime -}(1S,\frac{1}{2}^+)$\hfill$\gamma$ & 0.0 &0.0  &0.0  \\
$\Lambda_b^{0}(1P,\frac{3}{2}^-)$\hfill$\to$\hfill$\Sigma_b^{* 0}(1S,\frac{3}{2}^+)$\hfill$\gamma$ & 0.1 &0.15 &0.0  &$\Xi_b^{-}(1P,\frac{3}{2}^-)$\hfill$\to$\hfill$\Xi_b^{* -}(1S,\frac{3}{2}^+)$\hfill$\gamma$      & 0.0 &0.0  &0.0  \\
\Xcline{5-8}{0.75pt}
                                                                                                   &     &     &     &$\Xi_b^{0}(1P,\frac{1}{2}^-)$\hfill$\to$\hfill$\Xi_b^{0}(1S,\frac{1}{2}^+)$\hfill$\gamma$        &33.9 &63.6 &83.1 \\
                                                                                                   &     &     &     &$\Xi_b^{0}(1P,\frac{1}{2}^-)$\hfill$\to$\hfill$\Xi_b^{\prime 0}(1S,\frac{1}{2}^+)$\hfill$\gamma$ & 0.4 &1.32 &0.6  \\
                                                                                                   &     &     &     &$\Xi_b^{0}(1P,\frac{1}{2}^-)$\hfill$\to$\hfill$\Xi_b^{* 0}(1S,\frac{3}{2}^+)$\hfill$\gamma$      & 0.5 &2.04 &0.1  \\
                                                                                                   &     &     &     &$\Xi_b^{0}(1P,\frac{3}{2}^-)$\hfill$\to$\hfill$\Xi_b^{0}(1S,\frac{1}{2}^+)$\hfill$\gamma$        &35.2 &68.3 &88.9 \\
                                                                                                   &     &     &     &$\Xi_b^{0}(1P,\frac{3}{2}^-)$\hfill$\to$\hfill$\Xi_b^{\prime 0}(1S,\frac{1}{2}^+)$\hfill$\gamma$ & 0.6 &1.68 &0.7  \\
                                                                                                   &     &     &     &$\Xi_b^{0}(1P,\frac{3}{2}^-)$\hfill$\to$\hfill$\Xi_b^{* 0}(1S,\frac{3}{2}^+)$\hfill$\gamma$      & 0.6 &2.64 &0.2  \\
\end{tabular*}
}\\
\bottomrule[1.00pt]
\addlinespace[0.75em]
\toprule[1.00pt]
\multicolumn{10}{c}{
\begin{tabular*}{\textwidth}{@{\extracolsep{\fill}}lc@{\vrule width 0.75pt}lc@{\vrule width 0.75pt}lc}
\multicolumn{1}{c}{Process}&Our&\multicolumn{1}{c}{Process}&Our&\multicolumn{1}{c}{Process}&Our\\
\midrule[0.75pt]
$\Lambda_b^{0}(1D,\frac{3}{2}^+)$\hfill$\to$\hfill$\Lambda_b^{0}(1S,\frac{1}{2}^+)$\hfill$\gamma$  &26.7 &$\Xi_b^{-}(1D,\frac{3}{2}^+)$\hfill$\to$\hfill$\Xi_b^{-}(1S,\frac{1}{2}^+)$\hfill$\gamma$        &78.9 &$\Xi_b^{0}(1D,\frac{3}{2}^+)$\hfill$\to$\hfill$\Xi_b^{0}(1S,\frac{1}{2}^+)$\hfill$\gamma$        &10.2 \\
$\Lambda_b^{0}(1D,\frac{3}{2}^+)$\hfill$\to$\hfill$\Sigma_b^{0}(1S,\frac{1}{2}^+)$\hfill$\gamma$   & 1.1 &$\Xi_b^{-}(1D,\frac{3}{2}^+)$\hfill$\to$\hfill$\Xi_b^{\prime -}(1S,\frac{1}{2}^+)$\hfill$\gamma$ & 0.0 &$\Xi_b^{0}(1D,\frac{3}{2}^+)$\hfill$\to$\hfill$\Xi_b^{\prime 0}(1S,\frac{1}{2}^+)$\hfill$\gamma$ & 1.3 \\
$\Lambda_b^{0}(1D,\frac{3}{2}^+)$\hfill$\to$\hfill$\Sigma_b^{* 0}(1S,\frac{3}{2}^+)$\hfill$\gamma$ & 1.7 &$\Xi_b^{-}(1D,\frac{3}{2}^+)$\hfill$\to$\hfill$\Xi_b^{* -}(1S,\frac{3}{2}^+)$\hfill$\gamma$      & 0.0 &$\Xi_b^{0}(1D,\frac{3}{2}^+)$\hfill$\to$\hfill$\Xi_b^{* 0}(1S,\frac{3}{2}^+)$\hfill$\gamma$      & 2.1 \\
$\Lambda_b^{0}(1D,\frac{5}{2}^+)$\hfill$\to$\hfill$\Lambda_b^{0}(1S,\frac{1}{2}^+)$\hfill$\gamma$  &27.2 &$\Xi_b^{-}(1D,\frac{5}{2}^+)$\hfill$\to$\hfill$\Xi_b^{-}(1S,\frac{1}{2}^+)$\hfill$\gamma$        &80.0 &$\Xi_b^{0}(1D,\frac{5}{2}^+)$\hfill$\to$\hfill$\Xi_b^{0}(1S,\frac{1}{2}^+)$\hfill$\gamma$        &10.3 \\
$\Lambda_b^{0}(1D,\frac{5}{2}^+)$\hfill$\to$\hfill$\Sigma_b^{0}(1S,\frac{1}{2}^+)$\hfill$\gamma$   & 1.3 &$\Xi_b^{-}(1D,\frac{5}{2}^+)$\hfill$\to$\hfill$\Xi_b^{\prime -}(1S,\frac{1}{2}^+)$\hfill$\gamma$ & 0.0 &$\Xi_b^{0}(1D,\frac{5}{2}^+)$\hfill$\to$\hfill$\Xi_b^{\prime 0}(1S,\frac{1}{2}^+)$\hfill$\gamma$ & 1.4 \\
$\Lambda_b^{0}(1D,\frac{5}{2}^+)$\hfill$\to$\hfill$\Sigma_b^{* 0}(1S,\frac{3}{2}^+)$\hfill$\gamma$ & 1.9 &$\Xi_b^{-}(1D,\frac{5}{2}^+)$\hfill$\to$\hfill$\Xi_b^{* -}(1S,\frac{3}{2}^+)$\hfill$\gamma$      & 0.0 &$\Xi_b^{0}(1D,\frac{5}{2}^+)$\hfill$\to$\hfill$\Xi_b^{* 0}(1S,\frac{3}{2}^+)$\hfill$\gamma$      & 2.3 \\
$\Lambda_b^{0}(1F,\frac{5}{2}^-)$\hfill$\to$\hfill$\Lambda_b^{0}(1S,\frac{1}{2}^+)$\hfill$\gamma$  &15.5 &$\Xi_b^{-}(1F,\frac{5}{2}^-)$\hfill$\to$\hfill$\Xi_b^{-}(1S,\frac{1}{2}^+)$\hfill$\gamma$        &33.1 &$\Xi_b^{0}(1F,\frac{5}{2}^-)$\hfill$\to$\hfill$\Xi_b^{0}(1S,\frac{1}{2}^+)$\hfill$\gamma$        & 3.6 \\
$\Lambda_b^{0}(1F,\frac{5}{2}^-)$\hfill$\to$\hfill$\Sigma_b^{0}(1S,\frac{1}{2}^+)$\hfill$\gamma$   & 1.1 &$\Xi_b^{-}(1F,\frac{5}{2}^-)$\hfill$\to$\hfill$\Xi_b^{\prime -}(1S,\frac{1}{2}^+)$\hfill$\gamma$ & 0.0 &$\Xi_b^{0}(1F,\frac{5}{2}^-)$\hfill$\to$\hfill$\Xi_b^{\prime 0}(1S,\frac{1}{2}^+)$\hfill$\gamma$ & 0.7 \\
$\Lambda_b^{0}(1F,\frac{5}{2}^-)$\hfill$\to$\hfill$\Sigma_b^{* 0}(1S,\frac{3}{2}^+)$\hfill$\gamma$ & 2.0 &$\Xi_b^{-}(1F,\frac{5}{2}^-)$\hfill$\to$\hfill$\Xi_b^{* -}(1S,\frac{3}{2}^+)$\hfill$\gamma$      & 0.0 &$\Xi_b^{0}(1F,\frac{5}{2}^-)$\hfill$\to$\hfill$\Xi_b^{* 0}(1S,\frac{3}{2}^+)$\hfill$\gamma$      & 1.3 \\
$\Lambda_b^{0}(1F,\frac{7}{2}^-)$\hfill$\to$\hfill$\Lambda_b^{0}(1S,\frac{1}{2}^+)$\hfill$\gamma$  &15.4 &$\Xi_b^{-}(1F,\frac{7}{2}^-)$\hfill$\to$\hfill$\Xi_b^{-}(1S,\frac{1}{2}^+)$\hfill$\gamma$        &33.3 &$\Xi_b^{0}(1F,\frac{7}{2}^-)$\hfill$\to$\hfill$\Xi_b^{0}(1S,\frac{1}{2}^+)$\hfill$\gamma$        & 3.5 \\
$\Lambda_b^{0}(1F,\frac{7}{2}^-)$\hfill$\to$\hfill$\Sigma_b^{0}(1S,\frac{1}{2}^+)$\hfill$\gamma$   & 1.1 &$\Xi_b^{-}(1F,\frac{7}{2}^-)$\hfill$\to$\hfill$\Xi_b^{\prime -}(1S,\frac{1}{2}^+)$\hfill$\gamma$ & 0.0 &$\Xi_b^{0}(1F,\frac{7}{2}^-)$\hfill$\to$\hfill$\Xi_b^{\prime 0}(1S,\frac{1}{2}^+)$\hfill$\gamma$ & 0.7 \\
$\Lambda_b^{0}(1F,\frac{7}{2}^-)$\hfill$\to$\hfill$\Sigma_b^{* 0}(1S,\frac{3}{2}^+)$\hfill$\gamma$ & 2.0 &$\Xi_b^{-}(1F,\frac{7}{2}^-)$\hfill$\to$\hfill$\Xi_b^{* -}(1S,\frac{3}{2}^+)$\hfill$\gamma$      & 0.0 &$\Xi_b^{0}(1F,\frac{7}{2}^-)$\hfill$\to$\hfill$\Xi_b^{* 0}(1S,\frac{3}{2}^+)$\hfill$\gamma$      & 1.4 \\
$\Lambda_b^{0}(2S,\frac{1}{2}^+)$\hfill$\to$\hfill$\Lambda_b^{0}(1S,\frac{1}{2}^+)$\hfill$\gamma$  & 0.0 &$\Xi_b^{-}(2S,\frac{1}{2}^+)$\hfill$\to$\hfill$\Xi_b^{-}(1S,\frac{1}{2}^+)$\hfill$\gamma$        & 0.0 &$\Xi_b^{0}(2S,\frac{1}{2}^+)$\hfill$\to$\hfill$\Xi_b^{0}(1S,\frac{1}{2}^+)$\hfill$\gamma$        & 0.0 \\
$\Lambda_b^{0}(2S,\frac{1}{2}^+)$\hfill$\to$\hfill$\Sigma_b^{0}(1S,\frac{1}{2}^+)$\hfill$\gamma$   & 0.3 &$\Xi_b^{-}(2S,\frac{1}{2}^+)$\hfill$\to$\hfill$\Xi_b^{\prime -}(1S,\frac{1}{2}^+)$\hfill$\gamma$ & 0.0 &$\Xi_b^{0}(2S,\frac{1}{2}^+)$\hfill$\to$\hfill$\Xi_b^{\prime 0}(1S,\frac{1}{2}^+)$\hfill$\gamma$ & 0.0 \\
$\Lambda_b^{0}(2S,\frac{1}{2}^+)$\hfill$\to$\hfill$\Sigma_b^{* 0}(1S,\frac{3}{2}^+)$\hfill$\gamma$ & 2.1 &$\Xi_b^{-}(2S,\frac{1}{2}^+)$\hfill$\to$\hfill$\Xi_b^{* -}(1S,\frac{3}{2}^+)$\hfill$\gamma$      & 0.0 &$\Xi_b^{0}(2S,\frac{1}{2}^+)$\hfill$\to$\hfill$\Xi_b^{* 0}(1S,\frac{3}{2}^+)$\hfill$\gamma$      & 0.5 \\
\end{tabular*}
}\\
\bottomrule[1.00pt]
\bottomrule[1.00pt]
\end{tabular*}
\end{table*}

Among these decay channels of $\bar{3}_f$ single-charm baryons, we first consider the simplest situation, i.e., $\bar{3}_f\to 1S+\gamma$, where the $1S$ stands for the $\Lambda_c(1S)$, $\Sigma_c(1S)$, and $\Xi_c^{(\prime)}(1S)$ states. The numerical results are presented in Table~\ref{tab:bar3fto1S.c}.

For the $\Lambda_c^+(1P)$, the radiative decay widths of the $\Lambda_c^+(1S)/\Sigma_c(1S)\gamma$ modes are very small, where similar results are also obtained in Refs.~\cite{Wang:2017kfr,Ortiz-Pacheco:2023kjn}. Thus, the experimental detection of these decay channels experimentally are a real challenge. But for the $\Lambda_c^+(1D)$, the $\Lambda_c^+(1D,\frac{3}{2}^+)\to \Lambda_c^+(1S,\frac{1}{2}^+)\gamma$ and $\Lambda_c^+(1D,\frac{3}{2}^+)\to \Lambda_c^+(1S,\frac{1}{2}^+)\gamma$ channels have considerable decay widths. We therefore suggest that the experimental colleague could concentrate on the radiative decays of the $\Lambda_c^+(1D,\frac{3}{2}^+)$ and $\Lambda_c^+(1D,\frac{5}{2}^+)$. However, the radiative decay widths of the $\Lambda_c^+(1F)/\Lambda_c(2S)$ transitions into $\Lambda_c^+(1S)/\Sigma_c^+(1S)\gamma$ are very small, which shows that finding them is also a difficult task for experiments.

The $\Xi_c$ states contain two charge status, i.e., $\Xi_c^0$ and $\Xi_c^+$. In radiative decays, the $\Xi_c^0$ and $\Xi_c^+$ have significantly different properties. A typical example is the different decay widths of $\Xi_c^0(1P,\frac{1}{2}^-)/\Xi_c^0(1P,\frac{3}{2}^-)\to \Xi_c^0(1S,\frac{1}{2}^+)\gamma$ and $\Xi_c^+(1P,\frac{1}{2}^-)/\Xi_c^+(1P,\frac{3}{2}^-)\to \Xi_c^+(1S,\frac{1}{2}^+)\gamma$. For the $\Xi_c^0(1P,\frac{1}{2}^-)/\Xi_c^0(1P,\frac{3}{2}^-)\to \Xi_c^0(1S,\frac{1}{2}^+)\gamma$ transitions, the theoretical widths are about 200 keV, while the widths of $\Xi_c^+(1P,\frac{1}{2}^-)/\Xi_c^+(1P,\frac{3}{2}^-)\to \Xi_c^+(1S,\frac{1}{2}^+)\gamma$ are only of the order of a few keV. Similar results are also obtained in Refs.~\cite{Wang:2017kfr,Ortiz-Pacheco:2023kjn}. This could explain why the Belle Collaboration~\cite{Belle:2020ozq} successfully measured the relative branching ratios of $\Xi_c^0(1P,\frac{1}{2}^-)/\Xi_c^0(1P,\frac{3}{2}^-)\to \Xi_c^0(1S,\frac{1}{2}^+)\gamma$ with $\Xi_c^0(1P,\frac{1}{2}^-)/\Xi_c^0(1P,\frac{3}{2}^-)\to \Xi_c^{\prime+}(1S,\frac{1}{2}^+)\pi^-/\Xi_c^{*+}(1S,\frac{3}{2}^+)\pi^-\to \Xi_c^+(1S,\frac{1}{2}^+)\gamma\pi^-/\Xi_c^0(1S,\frac{1}{2}^+)\pi^+\pi^-$, but only gave upper branch ratio limits of $\Xi_c^+(1P,\frac{1}{2}^-)/\Xi_c^+(1P,\frac{3}{2}^-)\to \Xi_c^+(1S,\frac{1}{2}^+)\gamma$ with $\Xi_c^+(1P,\frac{1}{2}^-)/\Xi_c^+(1P,\frac{3}{2}^-)\to \Xi_c^{\prime0}(1S,\frac{1}{2}^+)\pi^+/\Xi_c^{*0}(1S,\frac{3}{2}^+)\pi^+\to \Xi_c^0(1S,\frac{1}{2}^+)\gamma\pi^+/\Xi_c^+(1S,\frac{1}{2}^+)\pi^-\pi^+$. On the other hand, if we treat the light flavor-spatial wave functions of $\Xi_c$ as antisymmetrical and $\Xi_c^\prime$ as symmetric, the couplings between the $\Xi_c^0$ and $ \Xi_c^{0\prime}\gamma$ are forbidden, since the the charge expectations satisfy $\langle e_1\rangle=\langle e_2\rangle=\langle e_3\rangle=0$. If we consider the small constituent quark mass gap of $m_d-m_s$, the decays are allowed. But, in general, the decay widths are suppressed, i.e., most of the calculated results are less than 0.1 keV as shown in Table~\ref{tab:bar3fto1S.c}. According to Table~\ref{tab:bar3fto1S.c} we also find that some decay channels can have significant radiative decay widths, including $\Xi_c^{0,+}(1D)\to \Xi_c^{0,+}(1S)\gamma$ and $\Xi_c^0(1F)\to \Xi_c^0(1S)\gamma$. However, for other radiative decay processes in Table~\ref{tab:bar3fto1S.c}, the calculated widths are about or less than 1 keV.

In Table~\ref{tab:bar3fto1S.c}, we observe that some decay channels have similar widths, such as $\Gamma_{\Xi_c^{0}(1P, \frac{1}{2}^-) \to \Xi_c^{0}(1S, \frac{1}{2}^+) \gamma} \approx \Gamma_{\Xi_c^{0}(1P, \frac{3}{2}^-) \to \Xi_c^{0}(1S, \frac{1}{2}^+) \gamma}$ and $\Gamma_{\Lambda_c^{+}(1D, \frac{3}{2}^+) \to \Lambda_c^{+}(1S, \frac{1}{2}^+) \gamma} \approx \Gamma_{\Lambda_c^{+}(1D, \frac{5}{2}^+) \to \Lambda_c^{+}(1S, \frac{1}{2}^+) \gamma}$. Further analysis suggests that these states can be categorized within the same $NL$ doublet. According to heavy quark spin symmetry, for the $\lambda$-mode orbital excited $\bar{3}_f$ states, they share the same $j\ell$ with $j_\ell = L$. Each doublet thus includes states with $J = |L - \frac{1}{2}|$ and $J = |L + \frac{1}{2}|$. To clearly present how the splittings affect the decay behavior, we expand the decay widths in terms of $\frac{1}{m_c}$. The numerical results are
\begin{widetext}
\begin{equation}\label{eq:charmexpandf}
\begin{split}
\Gamma_{\Lambda_c^{+}(1P,\frac{1}{2}^-) \to \Lambda_c^{+}(1S,\frac{1}{2}^+)}({\rm keV})\approx 0.41-0.82[\mathcal{O}(1/m_c)]+0.41[\mathcal{O}(1/m_c^2)],\\
\Gamma_{\Lambda_c^{+}(1P,\frac{3}{2}^-) \to \Lambda_c^{+}(1S,\frac{1}{2}^+)}({\rm keV})\approx 0.33+0.47[\mathcal{O}(1/m_c)]+0.67[\mathcal{O}(1/m_c^2)],
\end{split}
\end{equation}
\begin{equation}
\begin{split}
\Gamma_{\Lambda_c^{+}(1D,\frac{3}{2}^+) \to \Lambda_c^{+}(1S,\frac{1}{2}^+)}({\rm keV})\approx 39.55+3.15[\mathcal{O}(1/m_c)]+0.08[\mathcal{O}(1/m_c^2)],\\
\Gamma_{\Lambda_c^{+}(1D,\frac{5}{2}^+) \to \Lambda_c^{+}(1S,\frac{1}{2}^+)}({\rm keV})\approx 44.06-2.49[\mathcal{O}(1/m_c)]+0.11[\mathcal{O}(1/m_c^2)],
\end{split}
\end{equation}
\begin{equation}
\begin{split}
\Gamma_{\Lambda_c^{+}(1F,\frac{5}{2}^-) \to \Lambda_c^{+}(1S,\frac{1}{2}^+)}({\rm keV})\approx 4.76-0.35[\mathcal{O}(1/m_c)]+0.01[\mathcal{O}(1/m_c^2)],\\
\Gamma_{\Lambda_c^{+}(1F,\frac{7}{2}^-) \to \Lambda_c^{+}(1S,\frac{1}{2}^+)}({\rm keV})\approx 4.67+0.26[\mathcal{O}(1/m_c)]+0.01[\mathcal{O}(1/m_c^2)],
\end{split}
\end{equation}
\begin{equation}
\begin{split}
\Gamma_{\Xi_c^{0}(1P,\frac{1}{2}^-) \to \Xi_c^{0}(1S,\frac{1}{2}^+)}({\rm keV})\approx 205.43+22.32[\mathcal{O}(1/m_c)]+0.61[\mathcal{O}(1/m_c^2)],\\
\Gamma_{\Xi_c^{0}(1P,\frac{3}{2}^-) \to \Xi_c^{0}(1S,\frac{1}{2}^+)}({\rm keV})\approx 250.87-15.01[\mathcal{O}(1/m_c)]+0.90[\mathcal{O}(1/m_c^2)],
\end{split}
\end{equation}
\begin{equation}
\begin{split}
\Gamma_{\Xi_c^{0}(1D,\frac{3}{2}^+) \to \Xi_c^{0}(1S,\frac{1}{2}^+)}({\rm keV})\approx 19.29-2.83[\mathcal{O}(1/m_c)]+0.14[\mathcal{O}(1/m_c^2)],\\
\Gamma_{\Xi_c^{0}(1D,\frac{5}{2}^+) \to \Xi_c^{0}(1S,\frac{1}{2}^+)}({\rm keV})\approx 20.23+2.11[\mathcal{O}(1/m_c)]+0.17[\mathcal{O}(1/m_c^2)],
\end{split}
\end{equation}
\begin{equation}
\begin{split}
\Gamma_{\Xi_c^{0}(1F,\frac{5}{2}^-) \to \Xi_c^{0}(1S,\frac{1}{2}^+)}({\rm keV})\approx 27.70+1.29[\mathcal{O}(1/m_c)]+0.02[\mathcal{O}(1/m_c^2)],\\
\Gamma_{\Xi_c^{0}(1F,\frac{7}{2}^-) \to \Xi_c^{0}(1S,\frac{1}{2}^+)}({\rm keV})\approx 27.38-0.96[\mathcal{O}(1/m_c)]+0.02[\mathcal{O}(1/m_c^2)],
\end{split}
\end{equation}
\begin{equation}
\begin{split}
\Gamma_{\Xi_c^{+}(1P,\frac{1}{2}^-) \to \Xi_c^{+}(1S,\frac{1}{2}^+)}({\rm keV})\approx 0.77+1.37[\mathcal{O}(1/m_c)]+0.61[\mathcal{O}(1/m_c^2)],\\
\Gamma_{\Xi_c^{+}(1P,\frac{3}{2}^-) \to \Xi_c^{+}(1S,\frac{1}{2}^+)}({\rm keV})\approx 1.30-1.08[\mathcal{O}(1/m_c)]+0.90[\mathcal{O}(1/m_c^2)],
\end{split}
\end{equation}
\begin{equation}
\begin{split}
\Gamma_{\Xi_c^{+}(1D,\frac{3}{2}^+) \to \Xi_c^{+}(1S,\frac{1}{2}^+)}({\rm keV})\approx 26.36+3.31[\mathcal{O}(1/m_c)]+0.14[\mathcal{O}(1/m_c^2)],\\
\Gamma_{\Xi_c^{+}(1D,\frac{5}{2}^+) \to \Xi_c^{+}(1S,\frac{1}{2}^+)}({\rm keV})\approx 28.41-2.50[\mathcal{O}(1/m_c)]+0.17[\mathcal{O}(1/m_c^2)],
\end{split}
\end{equation}
\begin{equation}\label{eq:charmexpandl}
\begin{split}
\Gamma_{\Xi_c^{+}(1F,\frac{5}{2}^-) \to \Xi_c^{+}(1S,\frac{1}{2}^+)}({\rm keV})\approx 0.08-0.07[\mathcal{O}(1/m_c)]+0.02[\mathcal{O}(1/m_c^2)],\\
\Gamma_{\Xi_c^{+}(1F,\frac{7}{2}^-) \to \Xi_c^{+}(1S,\frac{1}{2}^+)}({\rm keV})\approx 0.07+0.05[\mathcal{O}(1/m_c)]+0.02[\mathcal{O}(1/m_c^2)].
\end{split}
\end{equation}
\end{widetext}
Here, $m_c$ is in unit of GeV. According to Eqs.~(\ref{eq:charmexpandf})$-$(\ref{eq:charmexpandl}), the leading-order widths within the same doublet have similar values. This suggests that heavy quark spin symmetry plays an important role in these processes. The next and next-next leading orders depend on $\frac{1}{m_c}$ and $\frac{1}{m_c^2}$. Notably, the coefficients of these terms exhibit significant differences between the $J=|L-\frac{1}{2}|$ and $J=|L+\frac{1}{2}|$ states. These differences primarily arise from the variations in the spatial wave functions and angular momentum structures.

For the $\bar{3}_f$ single-bottom baryons, we also calculate the radiative decay widths for the $\bar{3}_f \to 1S + \gamma$ processes. The numerical results are listed in Table~\ref{tab:bar3fto1S.b}.

In the $\Lambda_b$ states, $\Lambda_b(5912)$ and $\Lambda_b(5920)$ are good candidates for $\Lambda_c(1P, \frac{1}{2}^-)$ and $\Lambda_b(1P, \frac{3}{2}^-)$, respectively. However, these states do not have two-body OZI-allowed decay channels and are observed through the three-body strong decay channel $\Lambda_b^0 \pi^+ \pi^-$. In this context, they are considered narrow states, and their radiative decays may constitute considerable fractions of their total decay widths. Our calculations show that the widths for the $\Lambda_b^0 \gamma$ decays of $\Lambda_c(1P, \frac{1}{2}^-)$ and $\Lambda_b(1P, \frac{3}{2}^-)$ are approximately 40 keV, which is significantly larger than those for the $\Sigma_b^{(*)0} \gamma$ channels.

The $\Xi_b$ baryons have two charge states: negative and neutral, each with different electromagnetic properties. In the $\Xi_b$ family, the $\Xi_b^-(1P, \frac{3}{2}^-)$ candidate $\Xi_b^-(6100)$ was first observed by the CMS Collaboration and confirmed by the LHCb Collaboration. The $\Xi_b^0(1P, \frac{1}{2}^-)$ and $\Xi_b^0(1P, \frac{3}{2}^-)$ candidates were also established by the LHCb Collaboration. However, the $\Xi_b^-(1P, \frac{1}{2}^-)$ state has not been observed yet. As shown in Table~\ref{tab:bar3fto1S.b}, all $\Xi_b(1P) \to \Xi_b(1S) \gamma$ processes exhibit large widths. Therefore, besides searching for the $\Xi_b^-(1P, \frac{1}{2}^-)$ in the $\Xi_b^- \pi^+ \pi^-$ channel, we also suggest searching for it in the $\Xi_b^-(1S, \frac{1}{2}^+) \gamma$ channel.

The $\Lambda_b(1D)$ candidates, $\Lambda_b(6146)$ and $\Lambda_b(6152)$, and the $\Xi_b(1D)$ candidates, $\Xi_b(6327)$ and $\Xi_b(6333)$, were observed by the LHCb Collaboration. According to the measured widths, these are all narrow states. As shown in Table~\ref{tab:bar3fto1S.b}, the decay processes $\Lambda_b(1D) \to \Lambda_b(1S) \gamma$ and $\Xi_b(1D) \to \Xi_b(1S) \gamma$ also have considerable decay widths.

Until now, the candidates for $\Lambda_b(1F)$, $\Xi_b(1F)$, and $\Xi_b(2S)$ have not been observed. In our calculations, the predicted widths for $\Lambda_b^0(1F) \to \Lambda_b^0(1S)\gamma$ and $\Xi_b^-(1F) \to \Xi_b^-(1S)\gamma$ are approximately 15 and 33 keV, respectively. However, we do not find any significant decay widths for the $\Xi_b^0(1F)$, $\Lambda_b(2S)$, and $\Xi_b(2S)$ in the $1S\gamma$ channels.

To analyze the dependence of the decay widths on $\frac{1}{m_b}$, we expand the decay widths in terms of $\frac{1}{m_b}$, i.e.,
\begin{widetext}
\begin{equation}
\begin{split}
\Gamma_{\Lambda_b^{0}(1P,\frac{1}{2}^-) \to \Lambda_b^{0}(1S,\frac{1}{2}^+)}({\rm keV})\approx 46.65+1.94[\mathcal{O}(1/m_b)]+0.02[\mathcal{O}(1/m_b^2)],\\
\Gamma_{\Lambda_b^{0}(1P,\frac{3}{2}^-) \to \Lambda_b^{0}(1S,\frac{1}{2}^+)}({\rm keV})\approx 49.27-1.06[\mathcal{O}(1/m_b)]+0.02[\mathcal{O}(1/m_b^2)],
\end{split}
\end{equation}
\begin{equation}
\begin{split}
\Gamma_{\Lambda_b^{0}(1D,\frac{3}{2}^+) \to \Lambda_b^{0}(1S,\frac{1}{2}^+)}({\rm keV})\approx 26.77-0.25[\mathcal{O}(1/m_b)]+0.001[\mathcal{O}(1/m_b^2)],\\
\Gamma_{\Lambda_b^{0}(1D,\frac{5}{2}^+) \to \Lambda_b^{0}(1S,\frac{1}{2}^+)}({\rm keV})\approx 27.19+0.17[\mathcal{O}(1/m_b)]+0.001[\mathcal{O}(1/m_b^2)],
\end{split}
\end{equation}
\begin{equation}
\begin{split}
\Gamma_{\Lambda_b^{0}(1F,\frac{5}{2}^-) \to \Lambda_b^{0}(1S,\frac{1}{2}^+)}({\rm keV})\approx 15.45+0.03[\mathcal{O}(1/m_b)]+0.00002[\mathcal{O}(1/m_b^2)],\\
\Gamma_{\Lambda_b^{0}(1F,\frac{7}{2}^-) \to \Lambda_b^{0}(1S,\frac{1}{2}^+)}({\rm keV})\approx 15.44-0.02[\mathcal{O}(1/m_b)]+0.00002[\mathcal{O}(1/m_b^2)],
\end{split}
\end{equation}
\begin{equation}
\begin{split}
\Gamma_{\Xi_b^{-}(1P,\frac{1}{2}^-) \to \Xi_b^{-}(1S,\frac{1}{2}^+)}({\rm keV})\approx 79.63-2.72[\mathcal{O}(1/m_b)]+0.02[\mathcal{O}(1/m_b^2)],\\
\Gamma_{\Xi_b^{-}(1P,\frac{3}{2}^-) \to \Xi_b^{-}(1S,\frac{1}{2}^+)}({\rm keV})\approx 84.20+1.49[\mathcal{O}(1/m_b)]+0.03[\mathcal{O}(1/m_b^2)],
\end{split}
\end{equation}
\begin{equation}
\begin{split}
\Gamma_{\Xi_b^{-}(1D,\frac{3}{2}^+) \to \Xi_b^{-}(1S,\frac{1}{2}^+)}({\rm keV})\approx 78.78+0.54[\mathcal{O}(1/m_b)]+0.001[\mathcal{O}(1/m_b^2)],\\
\Gamma_{\Xi_b^{-}(1D,\frac{5}{2}^+) \to \Xi_b^{-}(1S,\frac{1}{2}^+)}({\rm keV})\approx 80.07-0.37[\mathcal{O}(1/m_b)]+0.001[\mathcal{O}(1/m_b^2)],
\end{split}
\end{equation}
\begin{equation}
\begin{split}
\Gamma_{\Xi_b^{-}(1F,\frac{5}{2}^-) \to \Xi_b^{-}(1S,\frac{1}{2}^+)}({\rm keV})\approx 33.11-0.06[\mathcal{O}(1/m_b)]+0.00004[\mathcal{O}(1/m_b^2)],\\
\Gamma_{\Xi_b^{-}(1F,\frac{7}{2}^-) \to \Xi_b^{-}(1S,\frac{1}{2}^+)}({\rm keV})\approx 33.24+0.05[\mathcal{O}(1/m_b)]+0.00004[\mathcal{O}(1/m_b^2)],
\end{split}
\end{equation}
\begin{equation}
\begin{split}
\Gamma_{\Xi_b^{0}(1P,\frac{1}{2}^-) \to \Xi_b^{0}(1S,\frac{1}{2}^+)}({\rm keV})\approx 33.57+1.76[\mathcal{O}(1/m_b)]+0.02[\mathcal{O}(1/m_b^2)],\\
\Gamma_{\Xi_b^{0}(1P,\frac{3}{2}^-) \to \Xi_b^{0}(1S,\frac{1}{2}^+)}({\rm keV})\approx 35.42-0.97[\mathcal{O}(1/m_b)]+0.03[\mathcal{O}(1/m_b^2)],
\end{split}
\end{equation}
\begin{equation}
\begin{split}
\Gamma_{\Xi_b^{0}(1D,\frac{3}{2}^+) \to \Xi_b^{0}(1S,\frac{1}{2}^+)}({\rm keV})\approx 10.28-0.20[\mathcal{O}(1/m_b)]+0.001[\mathcal{O}(1/m_b^2)],\\
\Gamma_{\Xi_b^{0}(1D,\frac{5}{2}^+) \to \Xi_b^{0}(1S,\frac{1}{2}^+)}({\rm keV})\approx 10.30+0.13[\mathcal{O}(1/m_b)]+0.001[\mathcal{O}(1/m_b^2)],
\end{split}
\end{equation}
\begin{equation}
\begin{split}
\Gamma_{\Xi_b^{0}(1F,\frac{5}{2}^-) \to \Xi_b^{0}(1S,\frac{1}{2}^+)}({\rm keV})\approx 3.55+0.02[\mathcal{O}(1/m_b)]+0.00004[\mathcal{O}(1/m_b^2)],\\
\Gamma_{\Xi_b^{0}(1F,\frac{7}{2}^-) \to \Xi_b^{0}(1S,\frac{1}{2}^+)}({\rm keV})\approx 3.52-0.01[\mathcal{O}(1/m_b)]+0.00004[\mathcal{O}(1/m_b^2)].
\end{split}
\end{equation}
\end{widetext}
Here, $m_b$ is in unit of GeV. By comparing our results with those for charmed baryons, we find that the decay widths within a doublet exhibit smaller differences in the bottom baryons. This can be attributed to the heavy quark spin symmetry, which is closely related to the mass of the heavy flavor quark. In the constituent quark model, we have $m_b > m_c$. Consequently, we expect that the radiative decay width gaps are smaller in bottom baryons.

\subsection{$6_f\to 1S+\gamma$}

\begin{table*}[htbp]
\caption{The radiative decay widths (in units of keV) of single-charm baryons in the $6_f$ representation decaying into the $\Lambda_c(1S)/\Sigma_c(1S)/\Xi_c^{(\prime)}(1S)/\Omega_c(1S)\gamma$.}
\label{tab:6fto1S.c}
\renewcommand\arraystretch{1.25}
\begin{tabular*}{\textwidth}{@{\extracolsep{\fill}}lcccccclcc}
\toprule[1.00pt]
\toprule[1.00pt]
\multicolumn{10}{c}{
\begin{tabular*}{\textwidth}{@{\extracolsep{\fill}}lcccclcccc}
\multicolumn{1}{c}{Process}		&Our &\makecell[c]{Ref.\\\cite{Wang:2017kfr}} &\makecell[c]{Ref.\\\cite{Ortiz-Pacheco:2023kjn}}  &\makecell[c]{Ref.\\\cite{Hazra:2021lpa}}                        &\multicolumn{1}{c}{Process}		&Our &\makecell[c]{Ref.\\\cite{Wang:2017kfr}} &\makecell[c]{Ref.\\\cite{Ortiz-Pacheco:2023kjn}}  &\makecell[c]{Ref.\\\cite{Hazra:2021lpa}}\\
\midrule[0.75pt]
$\Sigma_c^{* 0}(1S,\frac{3}{2}^+)$\hfill$\to$\hfill$\Sigma_c^{0}(1S,\frac{1}{2}^+)$\hfill$\gamma$   &  1.3 &3.43  &1.8   &1.378   &$\Xi_c^{\prime 0}(1S,\frac{1}{2}^+)$\hfill$\to$\hfill$\Xi_c^{0}(1S,\frac{1}{2}^+)$\hfill$\gamma$   & 0.3 &0.0   &0.4  &0.342 \\
$\Sigma_c^{+}(1S,\frac{1}{2}^+)$\hfill$\to$\hfill$\Lambda_c^{+}(1S,\frac{1}{2}^+)$\hfill$\gamma$    & 59.2 &80.6  &87.2  &93.5    &$\Xi_c^{* 0}(1S,\frac{3}{2}^+)$\hfill$\to$\hfill$\Xi_c^{0}(1S,\frac{1}{2}^+)$\hfill$\gamma$        & 1.1 &0.0   &1.6  &1.322 \\
$\Sigma_c^{* +}(1S,\frac{3}{2}^+)$\hfill$\to$\hfill$\Lambda_c^{+}(1S,\frac{1}{2}^+)$\hfill$\gamma$  &132.8 &373   &199.4 &231     &$\Xi_c^{* 0}(1S,\frac{3}{2}^+)$\hfill$\to$\hfill$\Xi_c^{\prime 0}(1S,\frac{1}{2}^+)$\hfill$\gamma$ & 1.0 &3.03  &1.4  &1.262 \\
$\Sigma_c^{* +}(1S,\frac{3}{2}^+)$\hfill$\to$\hfill$\Sigma_c^{+}(1S,\frac{1}{2}^+)$\hfill$\gamma$   &  0.0 &0.004 &0.0   &0.00067 &$\Xi_c^{\prime +}(1S,\frac{1}{2}^+)$\hfill$\to$\hfill$\Xi_c^{+}(1S,\frac{1}{2}^+)$\hfill$\gamma$   &14.9 &42.3  &20.6 &21.38 \\
$\Sigma_c^{* ++}(1S,\frac{3}{2}^+)$\hfill$\to$\hfill$\Sigma_c^{++}(1S,\frac{1}{2}^+)$\hfill$\gamma$ &  1.7 &3.94  &2.1   &1.483   &$\Xi_c^{* +}(1S,\frac{3}{2}^+)$\hfill$\to$\hfill$\Xi_c^{+}(1S,\frac{1}{2}^+)$\hfill$\gamma$        &52.7 &139   &74.2 &81.9  \\
                                                                                                    &      &      &      &        &$\Xi_c^{* +}(1S,\frac{3}{2}^+)$\hfill$\to$\hfill$\Xi_c^{\prime +}(1S,\frac{1}{2}^+)$\hfill$\gamma$ & 0.1 &0.004 &0.1  &0.029 \\
\Xcline{6-10}{0.75pt}
                                                                                                    &      &      &      &        &$\Omega_c^{* 0}(1S,\frac{3}{2}^+)$\hfill$\to$\hfill$\Omega_c^{0}(1S,\frac{1}{2}^+)$\hfill$\gamma$  & 0.9 &0.89  &1.0  &1.14  \\
\end{tabular*}
}\\
\bottomrule[1.00pt]
\addlinespace[0.75em]
\toprule[1.00pt]
\multicolumn{10}{c}{
\begin{tabular*}{\textwidth}{@{\extracolsep{\fill}}lccccccc}
&$\Sigma_c^0(1S,\frac{1}{2}^+)\gamma$ &$\Sigma_c^{*0}(1S,\frac{3}{2}^+)\gamma$&$\Lambda_c^+(1S,\frac{1}{2}^+)\gamma$ &$\Sigma_c^+(1S,\frac{1}{2}^+)\gamma$  &$\Sigma_c^{*+}(1S,\frac{3}{2}^+)\gamma$&$\Sigma_c^{++}(1S,\frac{1}{2}^+)\gamma$ &$\Sigma_c^{*++}(1S,\frac{3}{2}^+)\gamma$            \\
\midrule[0.75pt]
$\Sigma_{c0}^{0,+,++}(1P,\frac{1}{2}^-)$ &137.8 &183.0 &  0.0 & 3.3 & 2.7 &236.2 &281.6 \\
$\Sigma_{c1}^{0,+,++}(1P,\frac{1}{2}^-)$ &202.7 & 68.4 & 84.5 & 1.5 & 0.9 &279.0 & 93.4 \\
$\Sigma_{c1}^{0,+,++}(1P,\frac{3}{2}^-)$ & 59.3 &209.2 & 95.8 & 1.9 & 1.8 &101.1 &291.4 \\
$\Sigma_{c2}^{0,+,++}(1P,\frac{3}{2}^-)$ &158.8 & 25.5 & 49.9 & 1.9 & 0.7 &133.6 & 26.8 \\
$\Sigma_{c2}^{0,+,++}(1P,\frac{5}{2}^-)$ &  1.5 &148.2 & 53.5 & 1.1 & 0.7 & 11.1 &148.9 \\
$\Sigma_{c1}^{0,+,++}(1D,\frac{1}{2}^+)$ &  0.8 & 76.6 & 11.0 & 0.1 &56.0 &  2.5 &562.5 \\
$\Sigma_{c1}^{0,+,++}(1D,\frac{3}{2}^+)$ & 46.5 & 42.0 & 11.1 &32.3 &29.3 &330.8 &299.2 \\
$\Sigma_{c2}^{0,+,++}(1D,\frac{3}{2}^+)$ & 26.0 & 24.6 & 29.8 &22.9 &20.6 &214.6 &195.7 \\
$\Sigma_{c2}^{0,+,++}(1D,\frac{5}{2}^+)$ & 13.7 & 39.9 & 30.0 &10.0 &31.8 & 99.9 &308.8 \\
$\Sigma_{c3}^{0,+,++}(1D,\frac{5}{2}^+)$ & 13.6 &  4.9 & 14.6 &19.7 & 5.3 &156.6 & 45.7 \\
$\Sigma_{c3}^{0,+,++}(1D,\frac{7}{2}^+)$ &  1.1 & 19.5 & 14.6 & 0.1 &21.5 &  3.2 &186.8 \\
$\Sigma_{c2}^{0,+,++}(1F,\frac{3}{2}^-)$ &  0.2 & 63.5 &  3.7 & 0.1 & 7.3 &  1.2 &178.3 \\
$\Sigma_{c2}^{0,+,++}(1F,\frac{5}{2}^-)$ & 32.9 & 28.1 &  3.6 & 3.8 & 3.4 & 92.6 & 80.7 \\
$\Sigma_{c3}^{0,+,++}(1F,\frac{5}{2}^-)$ & 17.3 & 22.8 &  8.2 & 1.7 & 2.4 & 45.7 & 62.1 \\
$\Sigma_{c3}^{0,+,++}(1F,\frac{7}{2}^-)$ &  9.7 & 29.8 &  8.0 & 1.1 & 3.3 & 26.8 & 82.8 \\
$\Sigma_{c4}^{0,+,++}(1F,\frac{7}{2}^-)$ & 14.1 &  5.3 &  3.8 & 1.0 & 0.5 & 32.7 & 13.4 \\
$\Sigma_{c4}^{0,+,++}(1F,\frac{9}{2}^-)$ &  0.2 & 18.9 &  3.8 & 0.1 & 1.7 &  1.1 & 48.5 \\
$\Sigma_c^{0,+,++}(2S,\frac{1}{2}^+)$    & 14.6 &  0.4 &117.8 & 3.3 & 0.8 & 55.6 &  5.8 \\
$\Sigma_c^{* 0,+,++}(2S,\frac{3}{2}^+)$  & 10.8 & 11.5 &151.2 & 0.7 & 3.5 & 24.3 & 50.9 \\
\end{tabular*}
}\\
\bottomrule[1.00pt]
\addlinespace[0.75em]
\toprule[1.00pt]
\multicolumn{10}{c}{
\begin{tabular*}{\textwidth}{@{\extracolsep{\fill}}lcccccc@{\vrule width 0.75pt}lcc}
&$\Xi_c^0(1S,\frac{1}{2}^+)\gamma$ &$\Xi_c^{\prime0}(1S,\frac{1}{2}^+)\gamma$   &$\Xi_c^{*0}(1S,\frac{3}{2}^+)\gamma$&$\Xi_c^+(1S,\frac{1}{2}^+)\gamma$ &$\Xi_c^{\prime+}(1S,\frac{1}{2}^+)\gamma$   &$\Xi_c^{*+}(1S,\frac{3}{2}^+)\gamma$&&$\Omega_c^0(1S,\frac{1}{2}^+)\gamma$ &$\Omega_c^{*0}(1S,\frac{3}{2}^+)\gamma$\\
\midrule[0.75pt]
$\Xi_{c0}^{\prime 0,+}(1P,\frac{1}{2}^-)$ &0.0 &109.0 &150.1 & 0.0 & 1.2 & 0.6 &$\Omega_{c0}^{0}(1P,\frac{1}{2}^-)$ & 83.7 &110.2 \\
$\Xi_{c1}^{\prime 0,+}(1P,\frac{1}{2}^-)$ &0.5 &172.8 & 59.7 &36.3 & 0.0 & 1.0 &$\Omega_{c1}^{0}(1P,\frac{1}{2}^-)$ &143.6 & 46.9 \\
$\Xi_{c1}^{\prime 0,+}(1P,\frac{3}{2}^-)$ &0.6 & 50.1 &187.3 &44.9 & 1.6 & 0.3 &$\Omega_{c1}^{0}(1P,\frac{3}{2}^-)$ & 41.8 &154.3 \\
$\Xi_{c2}^{\prime 0,+}(1P,\frac{3}{2}^-)$ &0.3 &165.4 & 26.9 &24.3 & 9.6 & 2.2 &$\Omega_{c2}^{0}(1P,\frac{3}{2}^-)$ &168.3 & 26.2 \\
$\Xi_{c2}^{\prime 0,+}(1P,\frac{5}{2}^-)$ &0.4 &  0.7 &156.8 &27.1 & 1.7 & 5.1 &$\Omega_{c2}^{0}(1P,\frac{5}{2}^-)$ &  0.2 &155.1 \\
$\Xi_{c1}^{\prime 0,+}(1D,\frac{1}{2}^+)$ &0.0 &  0.4 & 32.4 & 4.4 & 0.1 &38.1 &$\Omega_{c1}^{0}(1D,\frac{1}{2}^+)$ &  0.2 & 11.4 \\
$\Xi_{c1}^{\prime 0,+}(1D,\frac{3}{2}^+)$ &0.0 & 20.9 & 18.4 & 4.6 &19.8 &19.5 &$\Omega_{c1}^{0}(1D,\frac{3}{2}^+)$ &  8.5 &  6.9 \\
$\Xi_{c2}^{\prime 0,+}(1D,\frac{3}{2}^+)$ &0.1 & 11.6 & 10.7 &12.5 &14.7 &14.4 &$\Omega_{c2}^{0}(1D,\frac{3}{2}^+)$ &  4.5 &  3.9 \\
$\Xi_{c2}^{\prime 0,+}(1D,\frac{5}{2}^+)$ &0.1 &  6.5 & 18.2 &12.9 & 6.1 &21.3 &$\Omega_{c2}^{0}(1D,\frac{5}{2}^+)$ &  2.8 &  7.2 \\
$\Xi_{c3}^{\prime 0,+}(1D,\frac{5}{2}^+)$ &0.0 &  5.9 &  2.3 & 6.5 &12.4 & 3.8 &$\Omega_{c3}^{0}(1D,\frac{5}{2}^+)$ &  2.1 &  0.9 \\
$\Xi_{c3}^{\prime 0,+}(1D,\frac{7}{2}^+)$ &0.0 &  0.6 &  9.3 & 6.5 & 0.1 &14.0 &$\Omega_{c3}^{0}(1D,\frac{7}{2}^+)$ &  0.4 &  3.9 \\
$\Xi_{c2}^{\prime 0,+}(1F,\frac{3}{2}^-)$ &0.0 &  0.1 & 39.3 & 1.3 & 0.1 & 0.2 &$\Omega_{c2}^{0}(1F,\frac{3}{2}^-)$ &  0.0 & 24.5 \\
$\Xi_{c2}^{\prime 0,+}(1F,\frac{5}{2}^-)$ &0.0 & 20.5 & 17.2 & 1.3 & 0.0 & 0.1 &$\Omega_{c2}^{0}(1F,\frac{5}{2}^-)$ & 13.0 & 10.7 \\
$\Xi_{c3}^{\prime 0,+}(1F,\frac{5}{2}^-)$ &0.0 & 12.0 & 15.5 & 3.0 & 0.1 & 0.1 &$\Omega_{c3}^{0}(1F,\frac{5}{2}^-)$ &  8.5 & 10.8 \\
$\Xi_{c3}^{\prime 0,+}(1F,\frac{7}{2}^-)$ &0.0 &  6.5 & 20.0 & 3.0 & 0.1 & 0.1 &$\Omega_{c3}^{0}(1F,\frac{7}{2}^-)$ &  4.5 & 13.8 \\
$\Xi_{c4}^{\prime 0,+}(1F,\frac{7}{2}^-)$ &0.0 & 11.7 &  4.2 & 1.4 & 0.3 & 0.1 &$\Omega_{c4}^{0}(1F,\frac{7}{2}^-)$ &  9.9 &  3.5 \\
$\Xi_{c4}^{\prime 0,+}(1F,\frac{9}{2}^-)$ &0.0 &  0.1 & 15.1 & 1.4 & 0.1 & 0.2 &$\Omega_{c4}^{0}(1F,\frac{9}{2}^-)$ &  0.0 & 12.3 \\
$\Xi_c^{\prime 0,+}(2S,\frac{1}{2}^+)$    &1.2 &  7.4 &  0.0 &46.9 & 6.2 & 1.1 &$\Omega_c^{0}(2S,\frac{1}{2}^+)$    &  3.0 &  0.2 \\
$\Xi_c^{* 0,+}(2S,\frac{3}{2}^+)$         &1.3 &  7.6 &  5.4 &74.9 & 1.4 & 6.0 &$\Omega_c^{* 0}(2S,\frac{3}{2}^+)$  &  5.4 &  1.9 \\
\end{tabular*}
}\\
\bottomrule[1.00pt]
\bottomrule[1.00pt]
\end{tabular*}
\end{table*}

\begin{table*}[htbp]
\caption{The radiative decay widths (in units of keV) of single-bottom baryons in the $6_f$ representation decaying into the $\Lambda_b(1S)/\Sigma_b(1S)/\Xi_b^{(\prime)}(1S)/\Omega_b(1S)\gamma$.}
\label{tab:6fto1S.b}
\renewcommand\arraystretch{1.25}
\begin{tabular*}{\textwidth}{@{\extracolsep{\fill}}lcccccclcc}
\toprule[1.00pt]
\toprule[1.00pt]
\multicolumn{10}{c}{
\begin{tabular*}{\textwidth}{@{\extracolsep{\fill}}lcccclcccc}
\multicolumn{1}{c}{Process}		&Our &\makecell[c]{Ref.\\\cite{Wang:2017kfr}} &\makecell[c]{Ref.\\\cite{Ortiz-Pacheco:2023kjn}}  &\makecell[c]{Ref.\\\cite{Hazra:2021lpa}}                        &\multicolumn{1}{c}{Process}		&Our &\makecell[c]{Ref.\\\cite{Wang:2017kfr}} &\makecell[c]{Ref.\\\cite{Ortiz-Pacheco:2023kjn}}  &\makecell[c]{Ref.\\\cite{Hazra:2021lpa}}\\
\midrule[0.75pt]
$\Sigma_b^{* -}(1S,\frac{3}{2}^+)$\hfill$\to$\hfill$\Sigma_b^{-}(1S,\frac{1}{2}^+)$\hfill$\gamma$  &  0.0 &0.06 &0.0   &0.0144 &$\Xi_b^{\prime -}(1S,\frac{1}{2}^+)$\hfill$\to$\hfill$\Xi_b^{-}(1S,\frac{1}{2}^+)$\hfill$\gamma$   & 0.6 &0.0  &0.6  &0.707  \\
$\Sigma_b^{0}(1S,\frac{1}{2}^+)$\hfill$\to$\hfill$\Lambda_b^{0}(1S,\frac{1}{2}^+)$\hfill$\gamma$   & 94.7 &130  &128.1 &151.9  &$\Xi_b^{* -}(1S,\frac{3}{2}^+)$\hfill$\to$\hfill$\Xi_b^{-}(1S,\frac{1}{2}^+)$\hfill$\gamma$        & 0.8 &0.0  &1.0  &1.044  \\
$\Sigma_b^{* 0}(1S,\frac{3}{2}^+)$\hfill$\to$\hfill$\Lambda_b^{0}(1S,\frac{1}{2}^+)$\hfill$\gamma$ &120.2 &335  &168.8 &198.8  &$\Xi_b^{* -}(1S,\frac{3}{2}^+)$\hfill$\to$\hfill$\Xi_b^{\prime -}(1S,\frac{1}{2}^+)$\hfill$\gamma$ & 0.0 &15.0 &0.0  &0.0122 \\
$\Sigma_b^{* 0}(1S,\frac{3}{2}^+)$\hfill$\to$\hfill$\Sigma_b^{0}(1S,\frac{1}{2}^+)$\hfill$\gamma$  &  0.0 &0.02 &0.0   &0.0059 &$\Xi_b^{\prime 0}(1S,\frac{1}{2}^+)$\hfill$\to$\hfill$\Xi_b^{0}(1S,\frac{1}{2}^+)$\hfill$\gamma$   &28.0 &84.6 &28.4 &46.9   \\
$\Sigma_b^{* +}(1S,\frac{3}{2}^+)$\hfill$\to$\hfill$\Sigma_b^{+}(1S,\frac{1}{2}^+)$\hfill$\gamma$  &  0.1 &0.25 &0.1   &0.080  &$\Xi_b^{* 0}(1S,\frac{3}{2}^+)$\hfill$\to$\hfill$\Xi_b^{0}(1S,\frac{1}{2}^+)$\hfill$\gamma$        &40.8 &104  &45.2 &65.0   \\
                                                                                                   &      &     &      &       &$\Xi_b^{* 0}(1S,\frac{3}{2}^+)$\hfill$\to$\hfill$\Xi_b^{\prime 0}(1S,\frac{1}{2}^+)$\hfill$\gamma$ & 0.0 &5.19 &0.0  &0.0069 \\
\Xcline{6-10}{0.75pt}
                                                                                                   &      &     &      &       &$\Omega_b^{* -}(1S,\frac{3}{2}^+)$\hfill$\to$\hfill$\Omega_b^{-}(1S,\frac{1}{2}^+)$\hfill$\gamma$  & 0.0 &0.1  &0.0  &0.056  \\
\end{tabular*}
}\\
\bottomrule[1.00pt]
\addlinespace[0.75em]
\toprule[1.00pt]
\multicolumn{10}{c}{
\begin{tabular*}{\textwidth}{@{\extracolsep{\fill}}lccccccc}
&$\Sigma_b^-(1S,\frac{1}{2}^+)\gamma$ &$\Sigma_b^{*-}(1S,\frac{3}{2}^+)\gamma$&$\Lambda_b^0(1S,\frac{1}{2}^+)\gamma$ &$\Sigma_b^0(1S,\frac{1}{2}^+)\gamma$  &$\Sigma_b^{*0}(1S,\frac{3}{2}^+)\gamma$&$\Sigma_b^{+}(1S,\frac{1}{2}^+)\gamma$ &$\Sigma_b^{*+}(1S,\frac{3}{2}^+)\gamma$            \\
\midrule[0.75pt]
$\Sigma_{b0}^{-,0,+}(1P,\frac{1}{2}^-)$ & 75.0 &128.2 &  0.0 &26.9 &46.7 &362.5 &624.6 \\
$\Sigma_{b1}^{-,0,+}(1P,\frac{1}{2}^-)$ &100.1 & 46.4 &116.8 &38.4 &17.6 &501.9 &230.7 \\
$\Sigma_{b1}^{-,0,+}(1P,\frac{3}{2}^-)$ & 30.0 &122.2 &122.5 &11.0 &46.7 &146.7 &610.7 \\
$\Sigma_{b2}^{-,0,+}(1P,\frac{3}{2}^-)$ & 56.2 & 12.3 & 63.7 &25.3 & 5.1 &308.0 & 64.3 \\
$\Sigma_{b2}^{-,0,+}(1P,\frac{5}{2}^-)$ &  1.7 & 66.2 & 66.0 & 0.4 &29.0 &  6.7 &357.4 \\
$\Sigma_{b1}^{-,0,+}(1D,\frac{1}{2}^+)$ &  0.9 &193.7 & 16.9 & 0.2 &44.3 &  3.6 &741.1 \\
$\Sigma_{b1}^{-,0,+}(1D,\frac{3}{2}^+)$ &102.2 &100.4 & 16.9 &23.4 &23.0 &391.0 &384.6 \\
$\Sigma_{b2}^{-,0,+}(1D,\frac{3}{2}^+)$ & 63.9 & 64.3 & 45.3 &14.4 &14.6 &243.0 &244.8 \\
$\Sigma_{b2}^{-,0,+}(1D,\frac{5}{2}^+)$ & 30.3 &100.1 & 45.4 & 6.9 &22.6 &115.6 &381.1 \\
$\Sigma_{b3}^{-,0,+}(1D,\frac{5}{2}^+)$ & 44.1 & 14.0 & 21.8 & 9.6 & 3.1 &165.1 & 52.7 \\
$\Sigma_{b3}^{-,0,+}(1D,\frac{7}{2}^+)$ &  1.0 & 57.4 & 21.9 & 0.3 &12.6 &  4.1 &215.6 \\
$\Sigma_{b2}^{-,0,+}(1F,\frac{3}{2}^-)$ &  0.5 & 88.7 &  6.2 & 0.1 &22.7 &  1.9 &359.2 \\
$\Sigma_{b2}^{-,0,+}(1F,\frac{5}{2}^-)$ & 47.1 & 39.9 &  6.0 &12.1 &10.2 &190.6 &161.7 \\
$\Sigma_{b3}^{-,0,+}(1F,\frac{5}{2}^-)$ & 23.6 & 31.4 & 13.8 & 6.1 & 8.0 & 95.7 &127.2 \\
$\Sigma_{b3}^{-,0,+}(1F,\frac{7}{2}^-)$ & 13.7 & 41.8 & 13.7 & 3.5 &10.7 & 55.5 &169.2 \\
$\Sigma_{b4}^{-,0,+}(1F,\frac{7}{2}^-)$ & 18.1 &  6.9 &  6.5 & 4.7 & 1.8 & 73.6 & 28.2 \\
$\Sigma_{b4}^{-,0,+}(1F,\frac{9}{2}^-)$ &  0.4 & 25.5 &  6.5 & 0.1 & 6.6 &  1.6 &103.4 \\
$\Sigma_b^{-,0,+}(2S,\frac{1}{2}^+)$    & 13.9 &  4.2 &172.2 & 3.5 & 0.9 & 55.7 & 15.8 \\
$\Sigma_b^{* -,0,+}(2S,\frac{3}{2}^+)$  &  4.8 & 15.9 &187.9 & 1.3 & 3.9 & 20.2 & 63.3 \\
\end{tabular*}
}\\
\bottomrule[1.00pt]
\addlinespace[0.75em]
\toprule[1.00pt]
\multicolumn{10}{c}{
\begin{tabular*}{\textwidth}{@{\extracolsep{\fill}}lcccccc@{\vrule width 0.75pt}lcc}
&$\Xi_b^-(1S,\frac{1}{2}^+)\gamma$ &$\Xi_b^{\prime-}(1S,\frac{1}{2}^+)\gamma$   &$\Xi_b^{*-}(1S,\frac{3}{2}^+)\gamma$&$\Xi_b^0(1S,\frac{1}{2}^+)\gamma$ &$\Xi_b^{\prime0}(1S,\frac{1}{2}^+)\gamma$   &$\Xi_b^{*0}(1S,\frac{3}{2}^+)\gamma$&&$\Omega_b^-(1S,\frac{1}{2}^+)\gamma$ &$\Omega_b^{*-}(1S,\frac{3}{2}^+)\gamma$\\
\midrule[0.75pt]
$\Xi_{b0}^{\prime -,0}(1P,\frac{1}{2}^-)$ &0.0 &44.5 & 75.5 & 0.0 &20.6 &35.6 &$\Omega_{b0}^{-}(1P,\frac{1}{2}^-)$ &30.9 &51.9 \\
$\Xi_{b1}^{\prime -,0}(1P,\frac{1}{2}^-)$ &0.7 &64.1 & 28.9 &44.9 &28.4 &13.4 &$\Omega_{b1}^{-}(1P,\frac{1}{2}^-)$ &48.3 &21.3 \\
$\Xi_{b1}^{\prime -,0}(1P,\frac{3}{2}^-)$ &0.7 &19.2 & 79.3 &49.3 & 8.5 &35.4 &$\Omega_{b1}^{-}(1P,\frac{3}{2}^-)$ &14.4 &60.2 \\
$\Xi_{b2}^{\prime -,0}(1P,\frac{3}{2}^-)$ &0.4 &43.9 &  9.0 &26.8 &16.8 & 3.9 &$\Omega_{b2}^{-}(1P,\frac{3}{2}^-)$ &39.9 & 7.8 \\
$\Xi_{b2}^{\prime -,0}(1P,\frac{5}{2}^-)$ &0.4 & 0.9 & 52.1 &28.6 & 0.6 &19.9 &$\Omega_{b2}^{-}(1P,\frac{5}{2}^-)$ & 0.6 &47.5 \\
$\Xi_{b1}^{\prime -,0}(1D,\frac{1}{2}^+)$ &0.0 & 0.3 &109.3 & 6.6 & 0.2 &18.4 &$\Omega_{b1}^{-}(1D,\frac{1}{2}^+)$ & 0.1 &70.0 \\
$\Xi_{b1}^{\prime -,0}(1D,\frac{3}{2}^+)$ &0.0 &58.8 & 57.3 & 6.7 & 9.4 & 9.6 &$\Omega_{b1}^{-}(1D,\frac{3}{2}^+)$ &38.2 &37.1 \\
$\Xi_{b2}^{\prime -,0}(1D,\frac{3}{2}^+)$ &0.1 &39.9 & 39.5 &18.0 & 5.3 & 5.8 &$\Omega_{b2}^{-}(1D,\frac{3}{2}^+)$ &28.4 &28.0 \\
$\Xi_{b2}^{\prime -,0}(1D,\frac{5}{2}^+)$ &0.1 &18.8 & 62.7 &18.4 & 2.7 & 8.7 &$\Omega_{b2}^{-}(1D,\frac{5}{2}^+)$ &13.3 &44.9 \\
$\Xi_{b3}^{\prime -,0}(1D,\frac{5}{2}^+)$ &0.1 &33.1 & 10.0 & 9.0 & 2.5 & 1.1 &$\Omega_{b3}^{-}(1D,\frac{5}{2}^+)$ &28.2 & 8.3 \\
$\Xi_{b3}^{\prime -,0}(1D,\frac{7}{2}^+)$ &0.1 & 0.5 & 43.0 & 9.1 & 0.2 & 3.6 &$\Omega_{b3}^{-}(1D,\frac{7}{2}^+)$ & 0.2 &36.7 \\
$\Xi_{b2}^{\prime -,0}(1F,\frac{3}{2}^-)$ &0.0 & 0.2 & 47.5 & 2.4 & 0.1 & 4.8 &$\Omega_{b2}^{-}(1F,\frac{3}{2}^-)$ & 0.1 &27.0 \\
$\Xi_{b2}^{\prime -,0}(1F,\frac{5}{2}^-)$ &0.0 &25.7 & 21.5 & 2.4 & 2.3 & 2.1 &$\Omega_{b2}^{-}(1F,\frac{5}{2}^-)$ &14.7 &12.3 \\
$\Xi_{b3}^{\prime -,0}(1F,\frac{5}{2}^-)$ &0.0 &13.6 & 17.8 & 5.3 & 1.1 & 1.6 &$\Omega_{b3}^{-}(1F,\frac{5}{2}^-)$ & 8.4 &11.1 \\
$\Xi_{b3}^{\prime -,0}(1F,\frac{7}{2}^-)$ &0.0 & 7.9 & 24.0 & 5.3 & 0.7 & 2.0 &$\Omega_{b3}^{-}(1F,\frac{7}{2}^-)$ & 4.9 &15.0 \\
$\Xi_{b4}^{\prime -,0}(1F,\frac{7}{2}^-)$ &0.0 &12.1 &  4.4 & 2.5 & 0.6 & 0.3 &$\Omega_{b4}^{-}(1F,\frac{7}{2}^-)$ & 8.7 & 3.1 \\
$\Xi_{b4}^{\prime -,0}(1F,\frac{9}{2}^-)$ &0.0 & 0.2 & 16.9 & 2.5 & 0.1 & 0.9 &$\Omega_{b4}^{-}(1F,\frac{9}{2}^-)$ & 0.1 &12.3 \\
$\Xi_b^{\prime -,0}(2S,\frac{1}{2}^+)$    &1.2 & 5.5 &  1.4 &64.4 & 4.6 & 1.1 &$\Omega_b^{-}(2S,\frac{1}{2}^+)$    & 2.3 & 0.4 \\
$\Xi_b^{* -,0}(2S,\frac{3}{2}^+)$         &1.3 & 2.2 &  6.4 &77.8 & 1.8 & 5.1 &$\Omega_b^{* -}(2S,\frac{3}{2}^+)$  & 1.1 & 2.8 \\
\end{tabular*}
}\\
\bottomrule[1.00pt]
\bottomrule[1.00pt]
\end{tabular*}
\end{table*}

In the radiative decay processes of $6_f$ single-charm baryons, we also calculate the widths of radiative decays with final states $\Lambda_c(1S)/\Sigma_c(1S)/\Xi_c^{(\prime)}(1S)/\Omega_c(1S)\gamma$ as the prior. The numerical results are given in Table~\ref{tab:6fto1S.c}.

The transitions $\Sigma_c^{(*)}(1S) \to \Lambda_c(1S)\gamma$ and $\Sigma_c^{*}(1S) \to \Sigma_c(1S)\gamma$ involve changes only in the spin wave functions, making them typical M1 transitions. In the radiative decays of $\Sigma_c(1S)$, the decay widths for $\Sigma_c^+(1S,\frac{1}{2}^+) \to \Lambda_c^+(1S,\frac{1}{2}^+)\gamma$ and $\Sigma_c^{*+}(1S,\frac{3}{2}^+) \to \Lambda_c^+(1S,\frac{1}{2}^+)\gamma$ are calculated to be 59.2 and 132.8 keV, respectively, consistent with the quark model calculations in Refs.~\cite{Wang:2017kfr,Ortiz-Pacheco:2023kjn,Hazra:2021lpa,Ivanov:1998wj,Ivanov:1999bk}, and also match the calculations with light cone QCD sum rules~\cite{Aliev:2009jt,Aliev:2016xvq,Zhu:1998ih}, heavy quark effective theory~\cite{Cheng:1992xi,Tawfiq:1999cf}, chiral perturbation theory~\cite{Jiang:2015xqa,Wang:2018cre}, and chiral quark-soliton model~\cite{Yang:2019tst,Kim:2021xpp}. Given that the $\Lambda_c^+(1S,\frac{1}{2}^+)$ is well established, these processes are promising candidates for experimental observation, which would be crucial for understanding the internal structure of baryons. The $\Sigma_c^0(1S,\frac{1}{2}^+)$ and $\Sigma_c^{++}(1S,\frac{1}{2}^+)$ states do not undergo radiative decays, as they are the lowest states with the same quark content. But $\Sigma_c^*(1S,\frac{3}{2}^+)$ can decay into $\Sigma_c(1S,\frac{1}{2}^+)\gamma$. The calculated widths of $\Sigma_c^{*0}(1S,\frac{3}{2}^+)$, $\Sigma_c^{*+}(1S,\frac{3}{2}^+)$, and $\Sigma_c^{*++}(1S,\frac{3}{2}^+)$ decaying into $\Sigma_c^0(1S,\frac{1}{2}^+)\gamma$, $\Sigma_c^+(1S,\frac{1}{2}^+)\gamma$, and $\Sigma_c^{++}(1S,\frac{1}{2}^+)\gamma$ are 1.3, $<0.1$, and 1.7 keV, respectively. Because of the small phase space, these widths are quite narrow. For the $\Sigma_c(1P)$ states, we note that some radiative decay widths are larger than 100 keV, such as $\Sigma_{c0}^0(1P,\frac{1}{2}^-)\to \Sigma_{c}^0(1S,\frac{1}{2}^+)\gamma$, $\Sigma_{c1}^0(1P,\frac{1}{2}^-)\to \Sigma_{c}^0(1S,\frac{1}{2}^+)\gamma$, and so on. But there are no large decay widths for the $\Sigma_c^+(1P)\to \Sigma_c^+(1S,\frac{1}{2}^+)\gamma$ and $\Sigma_c^+(1P)\to \Sigma_c^{*+}(1S,\frac{3}{2}^+)\gamma$ decays. According to Table~\ref{tab:6fto1S.c}, we suggest looking for the $\Sigma_c^0(1S,\frac{1}{2}^+)\gamma$, $\Sigma_c^{*0}(1S,\frac{3}{2}^+)\gamma$, $\Lambda_c^+(1S,\frac{1}{2}^+)\gamma$, $\Sigma_c^{++}(1S,\frac{1}{2}^+)\gamma$, and $\Sigma_c^{*++}(1S,\frac{3}{2}^+)\gamma$ modes of the $\Sigma_c(1P)$. For the the $\Sigma_c(1D)$ and $\Sigma_c(1F)$, we find large radiative decay widths from the $\Sigma_c^{++}(1S,\frac{1}{2}^+)\gamma$ and $\Sigma_c^{*++}(1S,\frac{3}{2}^+)\gamma$ channels. Finally, for the two $\Sigma_c(2S)$ states, the partial widths of the $\Lambda_c^+(1S,\frac{1}{2}^+)\gamma$ mode are calculated to be larger than 100 keV. In addition, 
the $\Sigma_c^{++}(1S,\frac{1}{2}^+)\gamma$ and $\Sigma_c^{*++}(1S,\frac{3}{2}^+)\gamma$ modes are also significant for the discussed $\Sigma_c(2S)$ state.

In the following, we will discuss the radiative decays of the $\Xi_c^\prime$ states. Two typical single-charm baryons are $\Xi_c^{\prime0}(1S,\frac{1}{2}^+)$ and $\Xi_c^{\prime+}(1S,\frac{1}{2}^+)$, where these two states have been observed in their radiative decays~\cite{CLEO:1998wvk}. In Sec.~\ref{subsec:bar3fto1S}, we point out that the coupling between the $\Xi_c^0$ state and $\Xi_c^{\prime0}\gamma$ is forbidden if we employ strict SU(3) flavor symmetry. The conclusion remains the same if we reverse the initial and final single-charm baryons in the above transition. But it is obviously in conflict with the experimental data~\cite{CLEO:1998wvk}. In Sec.~\ref{subsec:bar3fto1S}, we consider the mass differences of $m_d-m_s$ to obtain small decay widths. In this subsection, we use the same approach to obtain the decay width of $\Xi_c^{\prime0}(1S,\frac{1}{2}^+)\to\Xi_c^0(1S,\frac{1}{2}^+)\gamma$ at 0.3 keV. We propose to measure the width of the $\Xi_c^{\prime0}(1S,\frac{1}{2}^+)$ in high-precision experiments, which could test the SU(3) flavor symmetry and further unravel the structure of the single-charm baryon. In addition, the widths of the $\Xi_c^{\prime+}(1S,\frac{1}{2}^+)$ and $\Xi_c^{*+}(1S,\frac{3}{2}^+)$ decays into $\Xi_c^+(1S,\frac{1}{2}^+)\gamma$ are 14.9 and 52.7 keV, respectively. For the $\Xi_c^\prime(1P)$ states, we find some partial widths that are larger than 100 keV corresponding to the $\Xi_c^{\prime0}(1S,\frac{1}{2}^+)\gamma$ and $\Xi_c^{*0}(1S,\frac{3}{2}^+)\gamma$ channels. So far, there are several good $\Xi_c^{\prime}(1P)$ candidates that have been observed in experiment, i.e., the $\Xi_c(2880)$, $\Xi_c(2923)$, $\Xi_c(2939)$, and $\Xi_c(2965)$~\cite{LHCb:2020iby,LHCb:2022vns}. In previous experiments, these states have been studied by surveying their strong decays. In fact, the search for these states is also interesting using different approaches such as their radiative decays with $\Xi_c^{\prime0}(1S,\frac{1}{2}^+)\gamma$ and $\Xi_c^{*0}(1S,\frac{3}{2}^+)\gamma$ channels. In this work, we do not find the radiative decay widths of the discussed $\Xi_c^\prime(1D)$ and $\Xi_c^\prime(1F)$ states that are larger than 50 keV as indicated in Table~\ref{tab:6fto1S.c}. For the two $\Xi_c^{\prime}(2S)$ states, the decay widths of their $\Xi_c^{+}(1S,\frac{1}{2}^+)\gamma$ modes are much larger than other radiative decay channels.

For the $\Omega_c^{*0}(1S,\frac{3}{2}^+)$ state, the dominant radiative decay channel is $\Omega_c^0(1S,\frac{1}{2}^+)\gamma$, which has been observed in experiments~\cite{BaBar:2006pve,Solovieva:2008fw}. In 2017, the LHCb Collaboration~\cite{LHCb:2017uwr} observed the $\Omega_c(3000)$, $\Omega_c(3050)$, $\Omega_c(3065)$, $\Omega_c(3090)$, and $\Omega_c(3119)$ from the $pp$ collisions. According to the theoretical mass spectra and decay widths~\cite{Agaev:2017jyt,Agaev:2017lip,Ali:2017wsf,Chen:2017sci,Chen:2017gnu,Cheng:2017ove,Karliner:2017kfm,Padmanath:2017lng,Wang:2017hej,Wang:2017vnc,Wang:2017zjw,Yang:2020zrh,Yang:2017qan,Zhao:2017fov}, they could be considered as $\Omega_c(1P)$ or $\Omega_c(2S)$ candidates. In 2023, these states were confirmed in the same productions~\cite{LHCb:2023sxp}. In addition, except for $\Omega_c(3119)$, the remaining four states were observed in $e^+e^-$ collisions by the Belle Collaboration~\cite{Belle:2017ext} and in $\Omega_b^-\to \Omega_c^0(?) \pi^-\to \Xi_c^+K^-\pi^-$ by the LHCb Collaboration~\cite{LHCb:2021ptx}. However, their $J^P$ quantum numbers are still unknown. According to Table~\ref{tab:6fto1S.c}, the radiative decay channels $\Omega_c^0(1S,\frac{1}{2}^+)\gamma$ and $\Omega_c^{*0}(1S,\frac{3}{2}^+)\gamma$ of the $\Omega_c(1P)$ have larger widths. This gives us a new approach to study these states. We propose to study the $\Omega_c(1P)$ states in the $\Omega_c^0(1S,\frac{1}{2}^+)\gamma$ and $\Omega_c^{*0}(1S,\frac{3}{2}^+)\gamma$ channels. However, for the $\Omega_c(1D)$, $\Omega_c(1F)$, and $\Omega_c(2S)$, both partial widths of the $\Omega_c^0(1S,\frac{1}{2}^+)\gamma$ and $\Omega_c^{*0}(1S,\frac{3}{2}^+)\gamma$ channels are not obvious. Although observations of the higher excited states $\Omega_c(3188)$ and $\Omega_c(3327)$ help us to construct the $\Omega_c(2S)$ and $\Omega_c(1D)$ spectroscopy~\cite{LHCb:2017uwr,Belle:2017ext,LHCb:2023sxp,Yu:2023bxn,Luo:2023sra,Wang:2023wii,Karliner:2023okv,Pan:2023hwt}, the search for the $\Omega_c(1D)$, $\Omega_c(1F)$, and $\Omega_c(2S)$ in $\Omega_c^0(1S,\frac{1}{2}^+)\gamma$ and $\Omega_c^{*0}(1S,\frac{3}{2}^+)\gamma$ processes is really difficult in the expected future.

The radiative decays of $6_f$ single-bottom baryons into $\Lambda_b(1S) \gamma$, $\Sigma_b(1S) \gamma$, $\Xi_b^{(\prime)} \gamma$, and $\Omega_b(1S) \gamma$ are presented in Table~\ref{tab:6fto1S.b}. This table shows that there are no significant radiative decay widths for the $\Sigma_b^{\pm}$. For the $\Sigma_b^{(*)0}$, the decay widths involving the $\Lambda_b^0(1S, \frac{1}{2}^+)\gamma$ channel are approximately 100 keV. However, the neutral $\Sigma_c^{(*)0}$ states have not yet been observed. In addition to the $\Lambda_b^0\pi^0$ channel, the $\Lambda_b^0\gamma$ channel is also a viable method for searching for $\Sigma_c^{(*)0}$ states. Among the $\Xi_b^\prime(1S)$ states, the $\Xi_b^{\prime,-}(1S, \frac{1}{2}^+)$ and $\Xi_b^{*,0,-}(1S, \frac{3}{2}^+)$ have been established. However, the $\Xi_b^{\prime,0}(1S, \frac{1}{2}^+)$ remains undiscovered, possibly because the $\Xi_b^-\pi^+$ channel is kinematically forbidden. We note that the $\Xi_b^0\pi^0$ channel might be open, so we suggest searching for the $\Xi_b^{\prime,0}(1S, \frac{1}{2}^+)$ in the $\Xi_b^0\pi^0$ and $\Xi_b^0\gamma$ channels simultaneously. Finally, for the $\Omega_b^{*-}(1S, \frac{3}{2}^+)$, the calculated radiative decay widths are about or less than 0.1 keV. However, the $\Omega_b^{-}(1S, \frac{1}{2}^+)\gamma$ decay mode constitutes nearly 100\% of the branching fraction, making the observation of the $\Omega_b^{*-}(1S, \frac{3}{2}^+)$ via the $\Omega_b^{-}(1S, \frac{1}{2}^+)\gamma$ process a promising possibility.

Besides the $1S$ $6_f$ states, the LHCb Collaboration has recently observed some $1P$ candidates, namely $\Sigma_b(6097)$~\cite{LHCb:2018haf}, $\Xi_b(6227)$~\cite{LHCb:2018vuc}, $\Omega_b(6316)$, $\Omega_b(6330)$, $\Omega_b(6340)$, and $\Omega_b(6350)$~\cite{LHCb:2020tqd}. These states have been identified through their two-body strong decays in previous works. In this study, we also calculate their radiative decay widths. As shown in Table~\ref{tab:6fto1S.b}, some decay channels exhibit significantly large widths, such as 
$\Sigma_{b0}^+(1P, \frac{1}{2}^-)\to \Sigma_{b}^{*+}(1S, \frac{3}{2}^+)\gamma$ and $\Sigma_{b1}^+(1P, \frac{1}{2}^-)\to \Sigma_{b}^{+}(1S, \frac{1}{2}^+)\gamma$. 
These findings suggest that some missing $1P$ states may be discovered through their radiative decays.

Until now, there have been no strong candidates for the $1D$, $1F$, and $2S$ states in the $6_f$ bottom baryons. In this work, we calculate the mass spectra and radiative decay widths for these states. For the $\Sigma_b(1D)$ and $\Sigma_b(1F)$ states, we identify several radiative decay channels with widths in the hundreds of keV, which may provide valuable clues for searching for these states via radiative decays. However, for the $\Sigma_c(2S)$, $\Xi_c^{\prime}(1D)$, $\Xi_c^{\prime}(1F)$, $\Xi_c^{\prime}(2S)$, $\Omega_c(1D)$, $\Omega_c(1F)$, and $\Omega_c(2S)$ states, the radiative decay widths listed in Table~\ref{tab:6fto1S.b} are significantly smaller than those of the $\Sigma_b(1D)$ and $\Sigma_b(1F)$ states.

The calculations above are based on the quenched quark model, where the baryon is treated as a simple three-quark system. However, in the unquenched quark model, quark-antiquark pairs can be created from the vacuum, leading to coupled channel effects~\cite{Lu:2014ina,Garcia-Recio:2015jsa,Tornqvist:1984fy,Barnes:2007xu,Pennington:2007xr,Zhou:2011sp}. Coupled channel effects can influence the properties of hadrons. Previous studies~\cite{Isgur:1998kr,Geiger:1989yc} have suggested that if the mass of a hadron is near the thresholds of certain $S$-wave channels, significant coupled channel effects may arise. Notable examples include $D_{s0}^*(2317)$~\cite{vanBeveren:2003kd,Hwang:2004cd,Simonov:2004ar,Ortega:2016mms,Cheng:2017oqh,Dai:2003yg}, $D_{s1}^\prime(2460)$~\cite{Cheng:2017oqh,Dai:2003yg}, $X(3872)$~\cite{Li:2009ad,Kalashnikova:2005ui,Danilkin:2010cc,Duan:2020tsx}, and $\Lambda(1405)$~\cite{Silvestre-Brac:1991qqx}. The observed masses of these hadrons are approximately 100 MeV lower than the predictions of the quenched quark model, leading to what is known as the ``low mass puzzle." However, when the unquenched picture, including coupled channel effects, is considered~\cite{vanBeveren:2003kd,Hwang:2004cd,Simonov:2004ar,Ortega:2016mms,Cheng:2017oqh,Dai:2003yg,Li:2009ad,Kalashnikova:2005ui,Danilkin:2010cc,Duan:2020tsx,Silvestre-Brac:1991qqx}, this puzzle can be alleviated. The low mass puzzle is also evident in single-charm baryons. While most observed single-charm baryons align well with theoretical predictions from the quenched quark model, the $\Lambda_c(2940)$ stands out as a peculiar case. It is considered a candidate for the $\Lambda_c(2P)$ state; however, quenched quark models predict the mass of the $\Lambda_c(2P)$ to be about 60 to 100 MeV higher than the observed mass of the $\Lambda_c(2940)$. On the other hand, the $\Lambda_c(2940)$ is very close to the threshold of the $S$-wave $D^*N$ channel. In Refs.~\cite{Luo:2019qkm,Zhang:2022pxc}, the authors introduced the $D^*N$ channel in the $\Lambda_c(2P)$ calculation, and the mass was lowered to match the $\Lambda_c(2940)$. Given the significant coupled channel effect in $\Lambda_c(2940)$, its decay behaviors should also be studied within the unquenched quark model, considering the contributions of the $D^*N$ channel. However, the $\Lambda_c(2940)$ is an exceptional case. Although the conditions required for significant coupled-channel effects are stringent, they do not entirely rule out the existence of states similar to $\Lambda_c(2940)$. Currently, most of the observed low-lying singly heavy baryons align well with quenched quark model calculations~\cite{Ebert:2011kk,Yu:2022ymb,Yoshida:2015tia,Roberts:2007ni,Yamaguchi:2014era,Shah:2016nxi,Garcia-Tecocoatzi:2022zrf,Mao:2017wbz,Li:2022xtj}. However, several predicted $1P$, $1D$, and $2S$ states remain unobserved. In our previous work~\cite{Luo:2023sne}, we also calculated the mass spectra of $1F$ single-charm baryons, which have yet to be experimentally confirmed. As experimental techniques advance, we anticipate the discovery of a more complete family of singly heavy baryons. This will deepen our understanding of the applicability of the quenched quark model.

\subsection{Relativistic corrections in radiative decays}

In Eq.~(\ref{eq:he}), we consider only the two simplest terms of the Hamiltonian, representing the nonrelativistic approximation. However, in our calculations, the Hamiltonian involves more complex interactions~\cite{Li:1994cy,Li:1995si,Li:1993zzb,Brodsky:1968ea,LeYaouanc:1987gf}, necessitating the inclusion of relativistic corrections. It is important to discuss the model's applicability, but incorporating fully relativistic forms poses significant challenges. Therefore, we approximate by expanding the interactions in powers of $\frac{1}{m_q}$ or $\frac{1}{m_Q}$. Here, we take the relativistic correction~\cite{LeYaouanc:1987gf} term
\begin{equation}
h_e^{\rm RC}\sim\sum_j -e^{-i{\bm k}\cdot{\bm r}_j}\frac{{\bm \sigma}_j\times{\bm p}_j\cdot{\bm \epsilon}}{4m_j^2}.
\end{equation}
We now define the radiative decay width as follows:
\begin{equation}
\Gamma_{\rm RC}=\Gamma_{\rm NR}+\Delta\Gamma_{\rm RC},
\end{equation}
where $\Gamma_{\rm NR}$ represents the decay width obtained from the nonrelativistic model, where the Hamiltonian is given by Eq.~(\ref{eq:he}). The term $\Delta\Gamma_{\rm RC}$ denotes the relativistic correction, which quantifies the impact of relativistic effects on the radiative decay width. Below, we present examples to illustrate these corrections:
\begin{equation}\nonumber
\left(
\begin{array}{lcc}
{\rm Process}&{\makecell[c]{\Gamma_{\rm NR}}~{\rm (keV)}}&{\Delta\Gamma_{\rm RC}}~{\rm (keV)}\\
\mbox{$\Sigma_c^{*0 }(1S,\frac{3}{2}^+)\to \Sigma _c^{ 0}(1S,\frac{1}{2}^+)\gamma$} &  1.3 &\sim -0.07\\
\mbox{$\Sigma_c^{ + }(1S,\frac{1}{2}^+)\to \Lambda_c^{ +}(1S,\frac{1}{2}^+)\gamma$} & 59.2 &\sim -9.5\\
\mbox{$\Sigma_c^{*+ }(1S,\frac{3}{2}^+)\to \Lambda_c^{ +}(1S,\frac{1}{2}^+)\gamma$} &132.8 &\sim -27.5\\
\mbox{$\Sigma_c^{*+ }(1S,\frac{3}{2}^+)\to \Sigma _c^{ +}(1S,\frac{1}{2}^+)\gamma$} &  0.007 &\sim -0.002\\
\mbox{$\Sigma_c^{*++}(1S,\frac{3}{2}^+)\to \Sigma _c^{++}(1S,\frac{1}{2}^+)\gamma$} &  1.7 &\sim -0.15\\
\end{array}
\right),
\end{equation}
\begin{equation}\nonumber
\left(
\begin{array}{lcc}
{\rm Process}&{\makecell[c]{\Gamma_{\rm NR}}~{\rm (keV)}}&{\Delta\Gamma_{\rm RC}}~{\rm (keV)}\\
\mbox{$\Xi_c^{\prime 0}(1S,\frac{1}{2}^+)\to \Xi_c^{       0}(1S,\frac{1}{2}^+)\gamma$} &  0.3 &\sim -0.06\\
\mbox{$\Xi_c^{*      0}(1S,\frac{3}{2}^+)\to \Xi_c^{       0}(1S,\frac{1}{2}^+)\gamma$} &  1.1 &\sim -0.3\\
\mbox{$\Xi_c^{*      0}(1S,\frac{3}{2}^+)\to \Xi_c^{\prime 0}(1S,\frac{1}{2}^+)\gamma$} &  1.0 &\sim -0.04\\
\mbox{$\Xi_c^{\prime +}(1S,\frac{1}{2}^+)\to \Xi_c^{       +}(1S,\frac{1}{2}^+)\gamma$} & 14.9 &\sim -1.5\\
\mbox{$\Xi_c^{*      +}(1S,\frac{3}{2}^+)\to \Xi_c^{       +}(1S,\frac{1}{2}^+)\gamma$} & 52.7 &\sim -7.9\\
\mbox{$\Xi_c^{*      +}(1S,\frac{3}{2}^+)\to \Xi_c^{\prime +}(1S,\frac{1}{2}^+)\gamma$} &  0.1 &\sim -0.01\\
\end{array}
\right),
\end{equation}
\begin{equation}\nonumber
\left(
\begin{array}{lcc}
{\rm Process}&{\makecell[c]{\Gamma_{\rm NR}}~{\rm (keV)}}&{\Delta\Gamma_{\rm RC}}~{\rm (keV)}\\
\mbox{$\Omega_c^{*0}(1S,\frac{3}{2}^+)\to \Omega_c^{0}(1S,\frac{1}{2}^+)\gamma$} &  0.9 &\sim -0.03\\
\end{array}
\right),
\end{equation}
\begin{equation}\nonumber
\left(
\begin{array}{lcc}
{\rm Process}&{\makecell[c]{\Gamma_{\rm NR}}~{\rm (keV)}}&{\Delta\Gamma_{\rm RC}}~{\rm (keV)}\\
\mbox{$\Lambda_c^{+}(2S,\frac{1}{2}^+)\to \Lambda_c^{+}(1S,\frac{1}{2}^+)\gamma$} & 0.02  &\sim -0.02\\
\mbox{$\Lambda_c^{+}(1P,\frac{1}{2}^-)\to \Lambda_c^{+}(1S,\frac{1}{2}^+)\gamma$} & 0.1  &\sim 0.4\\
\mbox{$\Lambda_c^{+}(1D,\frac{3}{2}^+)\to \Lambda_c^{+}(1S,\frac{1}{2}^+)\gamma$} & 41.3  &\sim -1.5\\
\mbox{$\Lambda_c^{+}(1F,\frac{5}{2}^-)\to \Lambda_c^{+}(1S,\frac{1}{2}^+)\gamma$} &  4.6 &\sim 0.15\\
\end{array}
\right),
\end{equation}
\begin{equation}\nonumber
\left(
\begin{array}{lcc}
{\rm Process}&{\makecell[c]{\Gamma_{\rm NR}}~{\rm (keV)}}&{\Delta\Gamma_{\rm RC}}~{\rm (keV)}\\
\mbox{$\Xi_c^{0}(2S,\frac{1}{2}^+)\to \Xi_c^{0}(1S,\frac{1}{2}^+)\gamma$} & 0.04  &\sim -0.04\\
\mbox{$\Xi_c^{0}(1P,\frac{1}{2}^-)\to \Xi_c^{0}(1S,\frac{1}{2}^+)\gamma$} & 217.5  &\sim -13.7\\
\mbox{$\Xi_c^{0}(1D,\frac{3}{2}^+)\to \Xi_c^{0}(1S,\frac{1}{2}^+)\gamma$} &  17.8 &\sim 1.3\\
\mbox{$\Xi_c^{0}(1F,\frac{5}{2}^-)\to \Xi_c^{0}(1S,\frac{1}{2}^+)\gamma$} &  28.4 &\sim -0.6\\
\end{array}
\right).
\end{equation}
Based on the above calculations, we observe that some corrections are significant, such as in the cases of $\Sigma_c^{*+}(1S,\frac{3}{2}^+) \to \Lambda_c^+(1S,\frac{1}{2}^+)\gamma$ and $\Xi_c^0(1P,\frac{1}{2}^-) \to \Xi_c^0(1S,\frac{1}{2}^+)\gamma$. However, most corrections in the calculations are relatively small. We find no clear correlation between the magnitude of the corrections and the orbital or radial quantum numbers.

It is worth noting that ${\bm \sigma}$ and ${\bm p}$ are one-order irreducible tensors related to spin and spatial wave functions. One possible scenario is that when the spin or orbital quantum numbers change by one unit ($\Delta S=1$ or $\Delta L=1$) during the decay process, relativistic corrections might become significant. However, the calculations in this work are not fully relativistic. More serious discussions on this issue could be obtained using relativistic approaches such as the Bethe-Salpeter (BS) equations~\cite{Salpeter:1951sz,Salpeter:1952ib,Du:2024gna}, the light-front quark model~\cite{Ke:2011jf}, among others.

\section{Summary}\label{sec4}

At the beginning of the 21st century, the situation of observing singly heavy baryons has changed, as more and more singly heavy baryons have been reported in various experiments such as Belle and LHCb. It can be reflected by checking the presently collected singly heavy baryons in the PDG~\cite{ParticleDataGroup:2022pth}. Although the radiative decays of single-charm baryons have not been extensively reported compared to these observed strong and weak decay modes, the radiative decay of single-charm baryons is one aspect of the whole decay behavior of them that should be paid more attention. In particular, in the face of the high-precision era of hadorn spectroscopy, we have reason to believe that more radiative decays of singly heavy baryons will become experimentally accessible. With the promotion of experimental precision, we need to increase theoretical precision. 

In this work, we revisit the radiative decays of singly heavy baryons. In contrast to the treatment in the previous work~\cite{Luo:2023sne}, we adopt the numerical spatial wave function for the singly heavy baryon, which can be obtained by the concrete potential model and with that approach the GEM~\cite{Hiyama:2003cu}. This treatment can meet the requirements of high-precision hadron spectroscopy. We also make a comparison with the current experimental data~\cite{Belle:2020ozq} and the theoretical results from other groups~\cite{Wang:2017kfr,Ortiz-Pacheco:2023kjn,Hazra:2021lpa}, as summarized in the last section. 

With the running of Belle II~\cite{Kahn:2017ojr,Belle-II:2018jsg} and Run-3 and Run-4 of LHCb~\cite{DiNezza:2021rwt,Belin:2021jem}, we can expect more observations of singly heavy baryons. This information on the radiative decay of singly heavy baryons will stimulate our experimental colleagues to focus on exploring the spectroscopy of singly heavy baryons.

\begin{acknowledgments}
This work is supported by  the National Natural Science Foundation of China under Grants No.~12335001, No.~12247101, and No.~1240050206, National Key Research and Development Program of China under Contract No.~2020YFA0406400, the 111 Project under Grant No.~B20063, the fundamental Research Funds for the Central Universities, and the project for top-notch innovative talents of Gansu province.
\end{acknowledgments}

\appendix
\section{Some additional radiative decay widths}

In Tables~\ref{tab:bar3ftoext.c}$-$\ref{tab:6f1Ftoext.b}, we also list the calculated results of other radiative decay widths of singly heavy baryons. The final states of these discussed radiative decays are involved in radial or orbital excited states of singly heavy baryons. This information is left to the readers who are interested in exploring these allowed radiative decays.

\begin{table*}[htbp]
\caption{Radiative decay widths (in units of keV) of $3_f$ single-charm baryon decaying into radial or orbital excited states.}
\label{tab:bar3ftoext.c}
\renewcommand\arraystretch{1.25}
\begin{tabular*}{\textwidth}{@{\extracolsep{\fill}}lcccccccccccc}
\toprule[1.00pt]
\toprule[1.00pt]
&\multicolumn{3}{c}{$\Lambda_c^+/\Xi_c^{0,+}(1P,\frac{1}{2}^-)\gamma$}&\multicolumn{3}{c}{$\Lambda_c^+/\Xi_c^{0,+}(1P,\frac{3}{2}^-)\gamma$}&\multicolumn{3}{c}{$\Lambda_c^+/\Xi_c^{0,+}(2S,\frac{1}{2}^+)\gamma$}\\
\Xcline{2-4}{0.75pt}
\Xcline{5-7}{0.75pt}
\Xcline{8-10}{0.75pt}
                                            &$\Lambda_c^+$ &$\Xi_c^0$          &$\Xi_c^+$          &$\Lambda_c^+$ &$\Xi_c^0$          &$\Xi_c^+$          &$\Lambda_c^+$ &$\Xi_c^0$          &$\Xi_c^+$          &             &                   &                   \\
\midrule[0.75pt]
$\Lambda_c^+/\Xi_c^{0,+}(2S,\frac{1}{2}^+)$ &          0.3 &              45.9 &               0.0 &          0.2 &              68.1 &               0.1 &$\cdots$      &$\cdots$           &$\cdots$           &             &                   &                   \\
$\Lambda_c^+/\Xi_c^{0,+}(1D,\frac{3}{2}^+)$ &          0.3 &             270.4 &               1.2 &          0.1 &              46.0 &               0.4 &          0.1 &               0.0 &               0.0 &             &                   &                   \\
$\Lambda_c^+/\Xi_c^{0,+}(1D,\frac{5}{2}^+)$ &          0.3 &               0.6 &               0.1 &          1.1 &             298.8 &               0.6 &          0.3 &               0.1 &               0.1 &             &                   &                   \\
$\Lambda_c^+/\Xi_c^{0,+}(1F,\frac{5}{2}^-)$ &         74.3 &              35.7 &              55.2 &         19.8 &              10.0 &              15.6 &          1.1 &               3.7 &               0.2 &             &                   &                   \\
$\Lambda_c^+/\Xi_c^{0,+}(1F,\frac{7}{2}^-)$ &          0.7 &               0.7 &               0.2 &         81.5 &              48.4 &              61.7 &          1.2 &               3.9 &               0.3 &             &                   &                   \\
\end{tabular*}
\end{table*}

\begin{table*}[htbp]
\caption{Radiative decay widths (in units of keV) of $\Sigma_c(1P)/\Xi_c^{\prime}(1P)\Sigma_c(2S)/\Xi_c^{\prime}(2S)/\Omega_c(2S)$ decaying into radial or orbital excited states of the single-charm baryon.}
\renewcommand\arraystretch{1.25}
\begin{tabular*}{\textwidth}{@{\extracolsep{\fill}}lcccccccccccc}
\toprule[1.00pt]
\toprule[1.00pt]
&\multicolumn{3}{c}{$\Lambda_c^+/\Xi_c^{0,+}(1P,\frac{1}{2}^-)\gamma$}&\multicolumn{3}{c}{$\Lambda_c^+/\Xi_c^{0,+}(1P,\frac{3}{2}^-)\gamma$}&\multicolumn{3}{c}{$\Lambda_c^+/\Xi_c^{0,+}(2S,\frac{1}{2}^+)\gamma$}\\
\Xcline{2-4}{0.75pt}
\Xcline{5-7}{0.75pt}
\Xcline{8-10}{0.75pt}
                                                                                                      &$\Lambda_c^+$  &$\Xi_c^0$      &$\Xi_c^+$       &$\Lambda_c^+$     &$\Xi_c^0$         &$\Xi_c^+$      &$\Lambda_c^+$  &$\Xi_c^0$      &$\Xi_c^+$       &                  &                  &               \\
\midrule[0.75pt]
$\Sigma_{c0}^{+}/\Xi_{c0}^{\prime 0,+}(1P,\frac{1}{2}^-)$                                             &          28.4 &           0.2 &            7.8 &             33.0 &              0.2 &           7.3 &           0.0 &$\cdots$       &$\cdots$        &                  &                  &               \\
$\Sigma_{c1}^{+}/\Xi_{c1}^{\prime 0,+}(1P,\frac{1}{2}^-)$                                             &          43.7 &           0.2 &           11.5 &             12.2 &              0.1 &           2.5 &           0.0 &$\cdots$       &$\cdots$        &                  &                  &               \\
$\Sigma_{c1}^{+}/\Xi_{c1}^{\prime 0,+}(1P,\frac{3}{2}^-)$                                             &          15.3 &           0.1 &            5.5 &             47.7 &              0.3 &          14.8 &           0.0 &$\cdots$       &$\cdots$        &                  &                  &               \\
$\Sigma_{c2}^{+}/\Xi_{c2}^{\prime 0,+}(1P,\frac{3}{2}^-)$                                             &          61.3 &           0.5 &           22.2 &              7.2 &              0.0 &           2.3 &           0.0 &$\cdots$       &$\cdots$        &                  &                  &               \\
\end{tabular*}
\end{table*}

\begin{table*}[htbp]
\caption{Radiative decay widths (in units of keV) of $\Sigma_c(1D)/\Xi_c^{\prime}(1D)/\Omega_c(1D)$ decaying into radial or orbital excited states  of the single-charm baryon.}
\label{tab:6f1Dtoext.c}
\renewcommand\arraystretch{1.25}
\begin{tabular*}{\textwidth}{@{\extracolsep{\fill}}lcccccccccccc}
\toprule[1.00pt]
\toprule[1.00pt]
&\multicolumn{3}{c}{$\Lambda_c^+/\Xi_c^{0,+}(1P,\frac{1}{2}^-)\gamma$}&\multicolumn{3}{c}{$\Lambda_c^+/\Xi_c^{0,+}(1P,\frac{3}{2}^-)\gamma$}&\multicolumn{3}{c}{$\Lambda_c^+/\Xi_c^{0,+}(2S,\frac{1}{2}^+)\gamma$}\\
\Xcline{2-4}{0.75pt}
\Xcline{5-7}{0.75pt}
\Xcline{8-10}{0.75pt}
                                                                             &$\Lambda_c^+$  &$\Xi_c^0$      &$\Xi_c^+$       &$\Lambda_c^+$     &$\Xi_c^0$         &$\Xi_c^+$      &$\Lambda_c^+$  &$\Xi_c^0$      &$\Xi_c^+$       &                  &                  &               \\
\midrule[0.75pt]
$\Sigma_{c1}^{+}/\Xi_{c1}^{\prime 0,+}(1D,\frac{1}{2}^+)$                    &          38.7 &           0.2 &           16.1 &             19.7 &              0.1 &           7.6 &           0.4 &           0.0 &            0.0 &                  &                  &               \\
$\Sigma_{c1}^{+}/\Xi_{c1}^{\prime 0,+}(1D,\frac{3}{2}^+)$                    &          13.1 &           0.1 &            5.4 &             45.6 &              0.3 &          19.2 &           0.6 &           0.0 &            0.1 &                  &                  &               \\
$\Sigma_{c2}^{+}/\Xi_{c2}^{\prime 0,+}(1D,\frac{3}{2}^+)$                    &          86.4 &           0.6 &           38.1 &             20.4 &              0.1 &           8.7 &           1.0 &           0.0 &            0.1 &                  &                  &               \\
$\Sigma_{c2}^{+}/\Xi_{c2}^{\prime 0,+}(1D,\frac{5}{2}^+)$                    &           4.6 &           0.0 &            2.0 &             94.8 &              0.7 &          42.4 &           1.4 &           0.0 &            0.2 &                  &                  &               \\
\end{tabular*}
\end{table*}

\begin{table*}[htbp]
\caption{Radiative decay widths (in units of keV) of $\Sigma_c(1F)/\Xi_c^{\prime}(1F)/\Omega_c(1F)$ decaying into radial or orbital excited states of the single-charm baryon.}
\label{tab:6f1Ftoext.c}
\renewcommand\arraystretch{1.25}
\begin{tabular*}{\textwidth}{@{\extracolsep{\fill}}lcccccccccccc}
\toprule[1.00pt]
\toprule[1.00pt]
&\multicolumn{3}{c}{$\Lambda_c^+/\Xi_c^{0,+}(1P,\frac{1}{2}^-)\gamma$}&\multicolumn{3}{c}{$\Lambda_c^+/\Xi_c^{0,+}(1P,\frac{3}{2}^-)\gamma$}&\multicolumn{3}{c}{$\Lambda_c^+/\Xi_c^{0,+}(2S,\frac{1}{2}^+)\gamma$}\\
\Xcline{2-4}{0.75pt}
\Xcline{5-7}{0.75pt}
\Xcline{8-10}{0.75pt}
                                                                             &$\Lambda_c^+$  &$\Xi_c^0$      &$\Xi_c^+$       &$\Lambda_c^+$     &$\Xi_c^0$         &$\Xi_c^+$      &$\Lambda_c^+$  &$\Xi_c^0$      &$\Xi_c^+$       &                  &                  &               \\
\midrule[0.75pt]
$\Sigma_{c2}^{+}/\Xi_{c2}^{\prime 0,+}(1F,\frac{3}{2}^-)$                    &          23.3 &           0.1 &            9.5 &             10.3 &              0.0 &           3.9 &           2.0 &           0.0 &            0.3 &                  &                  &               \\
$\Sigma_{c2}^{+}/\Xi_{c2}^{\prime 0,+}(1F,\frac{5}{2}^-)$                    &           3.6 &           0.0 &            1.3 &             29.4 &              0.1 &          11.8 &           2.3 &           0.0 &            0.3 &                  &                  &               \\
$\Sigma_{c3}^{+}/\Xi_{c3}^{\prime 0,+}(1F,\frac{5}{2}^-)$                    &          45.9 &           0.1 &           19.1 &             14.6 &              0.0 &           5.8 &           3.6 &           0.0 &            0.6 &                  &                  &               \\
$\Sigma_{c3}^{+}/\Xi_{c3}^{\prime 0,+}(1F,\frac{7}{2}^-)$                    &           1.6 &           0.0 &            0.5 &             56.5 &              0.2 &          23.3 &           3.9 &           0.0 &            0.6 &                  &                  &               \\
\end{tabular*}
\end{table*}

\begin{table*}[htbp]
\renewcommand\arraystretch{1.25}
\begin{tabular*}{\textwidth}{@{\extracolsep{\fill}}lcccccccccccc}
{\it $-$Continue.$-$}\\
\toprule[1.00pt]
&\multicolumn{6}{c}{$\Sigma_{c}^{0,+,++}/\Xi_{c}^{\prime 0,+}/\Omega_{c}^0(2S,\frac{1}{2}^+)\gamma$}&\multicolumn{6}{c}{$\Sigma_{c}^{*0,+,++}/\Xi_{c}^{* 0,+}/\Omega_{c}^{*0}(2S,\frac{3}{2}^+)\gamma$}\\
\Xcline{2-7}{0.75pt}
\Xcline{8-13}{0.75pt}
                                                                             &$\Sigma_c^{0}$ &$\Sigma_c^{+}$ &$\Sigma_c^{++}$ &$\Xi_c^{\prime0}$ &$\Xi_c^{\prime+}$ &$\Omega_c^{0}$ &$\Sigma_c^{0}$ &$\Sigma_c^{+}$ &$\Sigma_c^{++}$ &$\Xi_c^{\prime0}$ &$\Xi_c^{\prime+}$ &$\Omega_c^{0}$ \\
\midrule[0.75pt]
$\Sigma_{c2}^{0,+,++}/\Xi_{c2}^{\prime 0,+}/\Omega_{c2}^0(1F,\frac{3}{2}^-)$ &           0.0 &           0.0 &            0.1 &              0.0 &              0.0 &           0.0 &           7.6 &           1.2 &           24.8 &              3.2 &              0.3 &           1.5 \\
$\Sigma_{c2}^{0,+,++}/\Xi_{c2}^{\prime 0,+}/\Omega_{c2}^0(1F,\frac{5}{2}^-)$ &           7.6 &           1.2 &           24.8 &              3.4 &              0.3 &           1.7 &           4.0 &           0.7 &           13.0 &              1.7 &              0.1 &           0.8 \\
$\Sigma_{c3}^{0,+,++}/\Xi_{c3}^{\prime 0,+}/\Omega_{c3}^0(1F,\frac{5}{2}^-)$ &           2.6 &           0.4 &            8.5 &              1.5 &              0.1 &           1.0 &           2.0 &           0.3 &            6.4 &              1.1 &              0.1 &           0.7 \\
$\Sigma_{c3}^{0,+,++}/\Xi_{c3}^{\prime 0,+}/\Omega_{c3}^0(1F,\frac{7}{2}^-)$ &           1.7 &           0.3 &            5.4 &              0.9 &              0.1 &           0.6 &           3.0 &           0.5 &            9.8 &              1.6 &              0.1 &           1.0 \\
\end{tabular*}
\end{table*}

\begin{table*}[htbp]
\caption{Radiative decay widths (in units of keV) of $3_f$ single-bottom baryon decaying into radial or orbital excited states.}
\label{tab:bar3ftoext.b}
\renewcommand\arraystretch{1.25}
\begin{tabular*}{\textwidth}{@{\extracolsep{\fill}}lcccccccccccc}
\toprule[1.00pt]
\toprule[1.00pt]
&\multicolumn{3}{c}{$\Lambda_b^0/\Xi_b^{-,0}(1P,\frac{1}{2}^-)\gamma$}&\multicolumn{3}{c}{$\Lambda_b^0/\Xi_b^{-,0}(1P,\frac{3}{2}^-)\gamma$}&\multicolumn{3}{c}{$\Lambda_b^0/\Xi_b^{-,0}(2S,\frac{1}{2}^+)\gamma$}\\
\Xcline{2-4}{0.75pt}
\Xcline{5-7}{0.75pt}
\Xcline{8-10}{0.75pt}
                                            &$\Lambda_b^0$ &$\Xi_b^-$          &$\Xi_b^0$          &$\Lambda_b^0$ &$\Xi_b^-$          &$\Xi_b^0$          &$\Lambda_b^0$ &$\Xi_b^-$          &$\Xi_b^0$          &             &                   &                   \\
\midrule[0.75pt]
$\Lambda_b^0/\Xi_b^{-,0}(2S,\frac{1}{2}^+)$ &         10.6 &              19.9 &               8.6 &         19.8 &              36.9 &              16.1 &$\cdots$      &$\cdots$           &$\cdots$           &             &                   &                   \\
$\Lambda_b^0/\Xi_b^{-,0}(1D,\frac{3}{2}^+)$ &         54.2 &             103.0 &              44.9 &         10.7 &              20.1 &               8.8 &          0.0 &               0.1 &               0.0 &             &                   &                   \\
$\Lambda_b^0/\Xi_b^{-,0}(1D,\frac{5}{2}^+)$ &          0.3 &               0.5 &               0.1 &         65.2 &             124.4 &              53.3 &          0.0 &               0.1 &               0.0 &             &                   &                   \\
$\Lambda_b^0/\Xi_b^{-,0}(1F,\frac{5}{2}^-)$ &         43.4 &             129.3 &              19.3 &         14.0 &              39.1 &               6.0 &          2.0 &               3.3 &               0.7 &             &                   &                   \\
$\Lambda_b^0/\Xi_b^{-,0}(1F,\frac{7}{2}^-)$ &          1.1 &               1.6 &               0.3 &         57.6 &             169.7 &              25.6 &          2.1 &               3.7 &               0.8 &             &                   &                   \\
\end{tabular*}
\end{table*}

\begin{table*}[htbp]
\caption{Radiative decay widths (in units of keV) of $\Sigma_b(1P)/\Xi_b^{\prime}(1P)\Sigma_b(2S)/\Xi_b^{\prime}(2S)/\Omega_b(2S)$ decaying into radial or orbital excited states of the single-bottom baryon.}
\renewcommand\arraystretch{1.25}
\begin{tabular*}{\textwidth}{@{\extracolsep{\fill}}lcccccccccccc}
\toprule[1.00pt]
\toprule[1.00pt]
&\multicolumn{3}{c}{$\Lambda_b^0/\Xi_b^{-,0}(1P,\frac{1}{2}^-)\gamma$}&\multicolumn{3}{c}{$\Lambda_b^0/\Xi_b^{-,0}(1P,\frac{3}{2}^-)\gamma$}&\multicolumn{3}{c}{$\Lambda_b^0/\Xi_b^{-,0}(2S,\frac{1}{2}^+)\gamma$}\\
\Xcline{2-4}{0.75pt}
\Xcline{5-7}{0.75pt}
\Xcline{8-10}{0.75pt}
                                                                                                     &$\Lambda_b^0$  &$\Xi_b^-$      &$\Xi_b^0$      &$\Lambda_b^0$     &$\Xi_b^-$         &$\Xi_b^0$      &$\Lambda_b^0$  &$\Xi_b^-$      &$\Xi_b^0$      &                  &                  &               \\
\midrule[0.75pt]
$\Sigma_{b0}^{0}/\Xi_{b0}^{\prime -,0}(1P,\frac{1}{2}^-)$                                            &          28.3 &           0.2 &           7.4 &             50.5 &              0.3 &          12.2 &           0.0 &$\cdots$       &$\cdots$       &                  &                  &               \\
$\Sigma_{b1}^{0}/\Xi_{b1}^{\prime -,0}(1P,\frac{1}{2}^-)$                                            &          45.6 &           0.3 &          12.5 &             20.5 &              0.1 &           5.1 &           0.0 &$\cdots$       &$\cdots$       &                  &                  &               \\
$\Sigma_{b1}^{0}/\Xi_{b1}^{\prime -,0}(1P,\frac{3}{2}^-)$                                            &          13.2 &           0.1 &           4.1 &             59.0 &              0.4 &          17.3 &           0.0 &$\cdots$       &$\cdots$       &                  &                  &               \\
$\Sigma_{b2}^{0}/\Xi_{b2}^{\prime -,0}(1P,\frac{3}{2}^-)$                                            &          53.6 &           0.4 &          17.3 &              9.6 &              0.1 &           2.9 &           0.0 &$\cdots$       &$\cdots$       &                  &                  &               \\
\end{tabular*}
\end{table*}

\begin{table*}[htbp]
\caption{Radiative decay widths (in units of keV) of $\Sigma_b(1D)/\Xi_b^{\prime}(1D)/\Omega_b(1D)$ decaying into radial or orbital excited states  of the single-bottom baryon.}
\label{tab:6f1Dtoext.b}
\renewcommand\arraystretch{1.25}
\begin{tabular*}{\textwidth}{@{\extracolsep{\fill}}lcccccccccccc}
\toprule[1.00pt]
\toprule[1.00pt]
&\multicolumn{3}{c}{$\Lambda_b^0/\Xi_b^{-,0}(1P,\frac{1}{2}^-)\gamma$}&\multicolumn{3}{c}{$\Lambda_b^0/\Xi_b^{-,0}(1P,\frac{3}{2}^-)\gamma$}&\multicolumn{3}{c}{$\Lambda_b^0/\Xi_b^{-,0}(2S,\frac{1}{2}^+)\gamma$}\\
\Xcline{2-4}{0.75pt}
\Xcline{5-7}{0.75pt}
\Xcline{8-10}{0.75pt}
                                                                            &$\Lambda_b^0$  &$\Xi_b^-$      &$\Xi_b^0$      &$\Lambda_b^0$     &$\Xi_b^-$         &$\Xi_b^0$      &$\Lambda_b^0$  &$\Xi_b^-$      &$\Xi_b^0$      &                  &                  &               \\
\midrule[0.75pt]
$\Sigma_{b1}^{-,0,+}/\Xi_{b1}^{\prime -,0}/\Omega_{b1}^-(1D,\frac{1}{2}^-)$ &          43.8 &           0.3 &          17.3 &             27.8 &              0.2 &           9.9 &           0.7 &           0.0 &           0.0 &                  &                  &               \\
$\Sigma_{b1}^{-,0,+}/\Xi_{b1}^{\prime -,0}/\Omega_{b1}^-(1D,\frac{3}{2}^-)$ &          15.0 &           0.1 &           5.6 &             58.5 &              0.3 &          22.8 &           0.8 &           0.0 &           0.1 &                  &                  &               \\
$\Sigma_{b2}^{-,0,+}/\Xi_{b2}^{\prime -,0}/\Omega_{b2}^-(1D,\frac{3}{2}^-)$ &          94.4 &           0.6 &          38.7 &             26.8 &              0.2 &          10.5 &           1.3 &           0.0 &           0.1 &                  &                  &               \\
$\Sigma_{b2}^{-,0,+}/\Xi_{b2}^{\prime -,0}/\Omega_{b2}^-(1D,\frac{5}{2}^-)$ &           5.1 &           0.0 &           2.0 &            117.7 &              0.8 &          48.4 &           1.6 &           0.0 &           0.1 &                  &                  &               \\
\end{tabular*}
\end{table*}

\begin{table*}[htbp]
\caption{Radiative decay widths (in units of keV) of $\Sigma_b(1F)/\Xi_b^{\prime}(1F)/\Omega_b(1F)$ decaying into radial or orbital excited states of the single-bottom baryon.}
\label{tab:6f1Ftoext.b}
\renewcommand\arraystretch{1.25}
\begin{tabular*}{\textwidth}{@{\extracolsep{\fill}}lcccccccccccc}
\toprule[1.00pt]
\toprule[1.00pt]
&\multicolumn{3}{c}{$\Lambda_b^0/\Xi_b^{-,0}(1P,\frac{1}{2}^-)\gamma$}&\multicolumn{3}{c}{$\Lambda_b^0/\Xi_b^{-,0}(1P,\frac{3}{2}^-)\gamma$}&\multicolumn{3}{c}{$\Lambda_b^0/\Xi_b^{-,0}(2S,\frac{1}{2}^+)\gamma$}\\
\Xcline{2-4}{0.75pt}
\Xcline{5-7}{0.75pt}
\Xcline{8-10}{0.75pt}
                                                                            &$\Lambda_b^0$  &$\Xi_b^-$      &$\Xi_b^0$      &$\Lambda_b^0$     &$\Xi_b^-$         &$\Xi_b^0$      &$\Lambda_b^0$  &$\Xi_b^-$      &$\Xi_b^0$      &                  &                  &               \\
\midrule[0.75pt]
$\Sigma_{b2}^{-,0,+}/\Xi_{b2}^{\prime -,0}/\Omega_{b2}^-(1F,\frac{3}{2}^-)$ &          29.6 &           0.1 &          12.2 &             15.3 &              0.0 &           5.5 &           3.3 &           0.0 &           0.4 &                  &                  &               \\
$\Sigma_{b2}^{-,0,+}/\Xi_{b2}^{\prime -,0}/\Omega_{b2}^-(1F,\frac{5}{2}^-)$ &           5.3 &           0.0 &           1.8 &             40.2 &              0.1 &          16.3 &           3.5 &           0.0 &           0.5 &                  &                  &               \\
$\Sigma_{b3}^{-,0,+}/\Xi_{b3}^{\prime -,0}/\Omega_{b3}^-(1F,\frac{5}{2}^-)$ &          58.1 &           0.2 &          23.9 &             21.1 &              0.1 &           7.9 &           5.3 &           0.0 &           0.8 &                  &                  &               \\
$\Sigma_{b3}^{-,0,+}/\Xi_{b3}^{\prime -,0}/\Omega_{b3}^-(1F,\frac{7}{2}^-)$ &           2.7 &           0.0 &           0.7 &             77.6 &              0.3 &          31.5 &           5.6 &           0.0 &           0.8 &                  &                  &               \\
\end{tabular*}
\end{table*}

\begin{table*}[htbp]
\renewcommand\arraystretch{1.25}
\begin{tabular*}{\textwidth}{@{\extracolsep{\fill}}lcccccccccccc}
{\it $-$Continue.$-$}\\
\toprule[1.00pt]
&\multicolumn{6}{c}{$\Sigma_{b}^{-,0,+}/\Xi_{b}^{\prime -,0}/\Omega_{b}^-(2S,\frac{1}{2}^+)\gamma$}&\multicolumn{6}{c}{$\Sigma_{b}^{*-,0,+}/\Xi_{b}^{* -,0}/\Omega_{b}^{*-}(2S,\frac{3}{2}^+)\gamma$}\\
\Xcline{2-7}{0.75pt}
\Xcline{8-13}{0.75pt}
                                                                            &$\Sigma_b^{-}$ &$\Sigma_b^{0}$ &$\Sigma_b^{+}$ &$\Xi_b^{\prime-}$ &$\Xi_b^{\prime0}$ &$\Omega_b^{-}$ &$\Sigma_b^{-}$ &$\Sigma_b^{0}$ &$\Sigma_b^{+}$ &$\Xi_b^{\prime-}$ &$\Xi_b^{\prime0}$ &$\Omega_b^{-}$ \\
\midrule[0.75pt]
$\Sigma_{b2}^{-,0,+}/\Xi_{b2}^{\prime -,0}/\Omega_{b2}^-(1F,\frac{3}{2}^-)$ &           0.0 &           0.0 &           0.1 &              0.0 &              0.0 &           0.0 &          12.5 &           3.2 &          50.2 &              4.6 &              1.1 &           1.8 \\
$\Sigma_{b2}^{-,0,+}/\Xi_{b2}^{\prime -,0}/\Omega_{b2}^-(1F,\frac{5}{2}^-)$ &           9.3 &           2.4 &          37.5 &              3.6 &              0.8 &           1.5 &           6.2 &           1.6 &          25.0 &              2.3 &              0.6 &           0.9 \\
$\Sigma_{b3}^{-,0,+}/\Xi_{b3}^{\prime -,0}/\Omega_{b3}^-(1F,\frac{5}{2}^-)$ &           2.9 &           0.7 &          11.7 &              1.4 &              0.3 &           0.7 &           2.9 &           0.7 &          11.7 &              1.4 &              0.3 &           0.7 \\
$\Sigma_{b3}^{-,0,+}/\Xi_{b3}^{\prime -,0}/\Omega_{b3}^-(1F,\frac{7}{2}^-)$ &           1.8 &           0.5 &           7.3 &              0.9 &              0.2 &           0.5 &           4.4 &           1.1 &          17.5 &              2.1 &              0.5 &           1.1 \\
\end{tabular*}
\end{table*}

\end{document}